\documentclass{aa}  
\usepackage{graphicx}
\usepackage{txfonts}

\newcommand{\Frac}[2]{\mbox{$\displaystyle\frac{#1}{#2}$}}
\newcommand{\Derpar}[2]{\frac{\partial #1}{\partial #2}}
\usepackage{amstext}

\begin{document}

        \title{Characterizing the astrometric precision
          limit\\ for moving targets observed with digital-array
          detectors\textsuperscript{*}}
          
          \titlerunning{Astrometric precision limit for moving targets}
          
        \author{S. Bouquillon\inst{1} \and
                        R. A. Mendez\inst{2} \and
                M. Altmann\inst{1,3} \and \\
                T. Carlucci\inst{1} \and 
                C. Barache\inst{1} \and 
                F. Taris\inst{1} \and           
                A.H. Andrei\inst{1,4} \and 
                R. Smart\inst{5}}
\institute{SYRTE, Observatoire de Paris, PSL Research University, CNRS, Sorbonne Universités, UPMC Univ. Paris 06, LNE, 61 avenue de l’Observatoire, 75014 Paris, France \\
                                \email{sebastien.bouquillon@obspm.fr}
         \and
                Departamento de Astronom\'\i a, 
                Facultad de Ciencias F\'\i sicas y Matem\' aticas, 
                Universidad de Chile, Casilla 36-D, Santiago, Chile
                \and
                                Zentrum f\" ur Astronomie der Universit\" at Heidelberg,
                                Astronomisches Recheninstitut,  
                                M\" onchhofstr. 12-14, 69120 Heidelberg, Germany
                \and
                            Observat\' orio Nacional, MCTI, Rua Gal. Jos\' e Cristino 77, 
                            Rio de Janeiro, RJ CEP 20921-400, Brasil
                \and
                                Instituto Nazionale di Astrofisica, 
                                Osservatorio Astrofisico di Torino, 
                                Strada Osservatorio 20, I-10025 Pino Torinese, Italy}

        \date{Received ... ; accepted ...}

        \abstract{} {We investigate the maximum astrometric precision
          that can be reached on moving targets observed with digital-sensor
          arrays, and provide an estimate for its ultimate lower
          limit based on the Cram\' er-Rao bound.}  {We extend
          previous work on one-dimensional Gaussian point-spread
functions (PSFs) focusing on
          moving objects and extending the scope to two-dimensional
          array detectors.  In this study the PSF of a stationary
          point-source celestial body is replaced by its convolution
          with a linear motion, thus effectively modeling the spread
          function of a moving target.}  {The expressions of the
          Cram\' er-Rao lower bound deduced by this method allow us
          to study in great detail the limit of astrometric precision
          that can be reached for moving celestial objects, and to compute an
          optimal exposure time according to different observational
          parameters such as seeing, detector pixel size, decentering,
          and elongation of the source caused by its drift. Comparison to 
          simulated and real data shows that the predictions of our 
          simple model are consistent with observations.}  {}


   \maketitle

\section{Introduction}
\label{sec:intro}  

{\let\thefootnote\relax\footnote{\textsuperscript{*} Based on data
    taken with the VST of the European Southern Observatory, programme
    092.B-0165 and 095.B-0046.}}
One of the crucial steps in obtaining accurate and
precise positions of objects on astronomical images is source
extraction and plate coordinate determination. The final astrometric
quality of the whole measurement process is dominated by this step.
Understanding the key mechanisms that define the precision of
an astrometric measurement is therefore paramount in order to be able to assess
the maximum precision that can be reached for a given detected source. This is 
particularly the case when the source in question is faint, which only
leads to images with a limited signal-to-noise ratio (S/N), or when the
requirements on the quality of the measurement are critical. For the
project described in this article, both cases apply.

The need for this study arose when we were preparing a campaign named GBOT 
to astrometrically observe the Gaia satellite from earthbound facilities,
a task that was required to ensure the full capabilities of Gaia
measurements, even for objects that have the most precise
measurements. For a description of the ``Ground Based Optical
Tracking'' project (GBOT), see, for example, \citet{Altmann14}. The
tight constraints on astrometric quality (i.e., precision and accuracy)
of 20~mas (1 mas = 1 milliarcsec) for a data point (on a daily basis)
led to the requirement of finding a centroiding mechanism as accurate and
as precise as possible for moving sources, and to analyze which
ultimate precision could be reached in theory.

Most of the astronomical projects involved in asteroid detection and
observation, such as Spacewatch \citep{Rabinowitz91} or Pan-STARRS
\citep{Kaiser10}, have used (and still largely use) the usual
two-dimensional Gaussian as the point-spread function (PSF) of moving
objects to detect the asteroids and to photometrically and
astrometrically reduce them. A two-dimensional Gaussian is not
well suited to represent the PSF of moving objects,
especially when the speed of the target is high. At the same time, finding programs for fast-moving objects (such as near-Earth
objects, NEOs), more specific
PSFs or centroiding methods have been proposed (see, for instance,
\citet{Kouprianov08} or \citet{Mao08}), but these methods are really
specific and cannot be used for slowly moving targets.

As we need a PSF model that can be applied to any moving source
regardless of its speed, we have developed the moving-Gaussian approach,
which was found to be most promising. Since a very similar technique
was independently developed and tested for Pan-STARRS purposes \citep{Veres12} (they call it ``trail fitting''), we cannot claim
generic authorship of this technique and therefore only
present a brief synopsis including the expressions required for our
study.

The incentive to rigorously analyze the limits of the informatory content
of astrometric signals was initially caused by the finding that the
target of the GBOT campaign (the Gaia satellite) was found to be much
fainter than expected, by no less than a full three magnitudes. This
meant completely reassessing our strategies, and before
doing that, we needed to estimate whether the aims of GBOT in terms of astrometric
precision would even be reachable. For this we urgently required a
theoretical foundation, which we found in the Cram\'er-Rao lower bound
(CRLB) analysis, as conducted by Mendez et al. (2013, MSL13
hereafter).  Following the path of MSL13 and \citet{Mendez14}, we
subjected our centroiding method to an analysis of the CRLB. While the
first paper, MSL13, explores the case of a one-dimensional Gaussian
PSF and focuses on the astrometric aspects alone, and the second paper,
\citet{Mendez14}, includes photometry in the same analysis, we extend
these works to a Gaussian PSF of a moving object (moving Gaussian,
hereafter called MoG), and give an analysis of the CRLB for the one
dimensional case, similar to the earlier studies, and extend this to
the full two-dimensional case. For an asymmetrical PSF, such as the
MoG, extending our analysis to the full two dimensions in order to
appreciate the theoretical precision limits as characterized by the
CRLB is much more significant than in the case of the circularly
symmetric Gaussian PSF analyzed previously.

Our theoretical results were then compared with observational data as they are
routinely derived in the course of our GBOT astrometric tracking
campaign. The bulk of these data has been obtained with ESO's VST, a
2.6~m telescope equipped with the $8\times 4\times 2048 \times 4096$
pixel OmegaCam mosaic array. This telescope tracks the moving
target, which means that the background stars have a trailed
PSF. Therefore, we are in the fortunate situation to have at
our disposal objects as input parameter
that encompass a wide range of apparent brightness, thus a large
coverage of S/N. Moreover, the
speed of Gaia is variable, resulting in a range of input trail
lengths.  This allows us to access a significant portion of the
possible parameter space that goes into the CRLB analysis - and can
thus substantiate our theoretical findings with observational proof
for most scenarios an observer would face under realistic conditions,
see Sect.~\ref{sec:CompWithImages}.

The results presented here are not only significant to the initial
question concerning the feasibility of our Gaia tracking program
given the unexpected faintness of the target, but can also be used to
estimate the requirements when planning and developing similar
enterprises. Moreover, they can give valuable estimates of the
precision of asteroid astrometry, aiding in the kinematic studies of
these objects, especially in high-precision measurements, for
example, when
trying to determine the magnitude of the non-gravitation motion of small
solar system bodies, such as that caused by the Yarkovsky effect, see, for instance, \citet{Nugent12}.

Section~\ref{sec:SpreadFun} introduces the spread function of point-like
moving objects for one- and two-dimensional array detectors.
Section~\ref{sec:StudyCR_1D} presents our study of the astrometric CRLB 
for a moving source observed with a one-dimensional
array detector. We note that even though linear detectors are rarely used for astronomical purposes, introducing the CRLB framework and equations for this simple case allows a much easier understanding of the relevance of these statistical tools as well as the main features of the astrometric behavior of moving objects observed with digital sensors.  Section~\ref{sec:CR_moving_2D} extends the studies of
the previous section to the case of a standard two-dimensional CCD
sensor for stationary sources (a result not yet published, to the best of our knowledge) as well as for moving sources, and analyzes the
differences with the one-dimensional array detector case.  Finally,
Sect.~\ref{sec:CompWithImages} compares the main results of this paper
with the astrometric precision of simulated and real astronomical
observations of moving sources that are observed with a two-dimensional sensor.

\section{Spread function of moving point sources}  
\label{sec:SpreadFun} 
Throughout this paper we make the simplifying assumption that
at each instant $t$, the flux of
photons arising from a moving or non-moving point-like object and
reaching the digital sensor follows a circularly symmetric Gaussian
distribution of light. We also assume that of the parameters of
this instantaneous distribution of photons, only the position of its
center can change during the exposure time $T_e$.  
These assumptions concerning the PSF model are necessary to work on
the resolution of CRLB equations in a generic framework, but
we show in Sect.~\ref{sec:CompWithImages} that the predictions of these equations are quite consistent with the results from real observations.
In particular, this means that the standard deviation $\sigma$, identical in all directions, and the total expected flux of photons $f_s$ received from the source per unit of time and reaching the CCD sensor, are both constant during
the whole exposure time.  This implies that over this short timescale, there is no variation in the brightness of the source,
and no variation of the sky or instrumental conditions (this also
implies that the source motion remains contained in a small area of
the CCD to avoid any instrumental aberration). 
We note, however, that we do include shot noise on the source and background and read-out-noise from the detector in our analysis.  

In this paper, the sky
and instrumental conditions are characterized through the full-width at half-maximum (FWHM hereafter) of a non-moving object. In the
case of the moving object, the motion of the point-like moving object
is assumed to be linear during the exposure time, which implies that
there is no rotation of the camera field of view, and no acceleration
of the motion of the object (or at least that these effects are
negligible over the time of exposure).

For a stationary source, the instantaneous distribution of
photons is therefore independent of time, and the total PSF resulting from the
integration over the exposure time is still a circularly symmetric
Gaussian, with unchanged standard deviation and center position.
Then, hereafter, the expressions of the total PSF after an exposure
time $T_e$ are given by $\Phi_S$ for stationary sources
observed with a one-dimensional detector, and 
by $\Phi_{S^2}$ for stationary sources observed with a
two-dimensional detector,
\begin{eqnarray}
\Phi_S \left( x-x_c\right)
&=&\tilde{F}~\overline{\Phi}_S \left( x-x_c\right)\label{eqS}\\ 
\Phi_{S^2} \left( x-x_c,y-y_c\right)&=&\tilde{F} ~ \overline{\Phi}_{S^2}\left( x-x_c,y-y_c\right)\nonumber\\
&=&\tilde{F} ~ \overline{\Phi}_S \left( x-x_c\right)~ \overline{\Phi}_S \left( y-y_c\right) 
\label{eqS-2D}
,\end{eqnarray}
where 
\begin{eqnarray}
\overline{\Phi}_S \left( z\right)&=&\frac{1}{\sqrt{2\pi}\sigma}~ e^{-\frac{1}{2} \left( \frac{z}{\sigma}\right)^2}\nonumber
\end{eqnarray}
$\overline{\text{\text{}}\Phi}_S\left( z\right)$ is the normalized one-dimensional
Gaussian function centered at zero where $z = x-x_c$ or $z = y-y_c$
depending on the case, and $\overline{\Phi}_{S^2}$ is the normalized circularly symmetric two-dimensional Gaussian.
In these expressions, $\tilde{F}$ is the source total flux (in
photon-$e^-$) with $\tilde{F} = f_s~T_e$, and $\sigma$ is the
standard deviation measuring the atmospheric and instrumental
scattering level. We note that FWHM $= 2\sqrt{2\ln{2}}\sigma$. 
For a linear sensor, $x$ is the coordinate along the axes of pixels,
while $x_c$ is the coordinate of the PSF center. For a two-dimensional
sensor, the coordinate system $(x,y)$ is the usual right-handed
orthonormal coordinate system with its origin at the bottom left
corner of the CCD, the $x$-coordinate along the bottom side of the CCD,
and the $y$-coordinate along the left side of the CCD. The coordinates
$(x_c,y_c)$ are the position of the PSF center in this reference
frame.


 
In contrast to the stationary case, the instantaneous distribution of
photons of a moving object is time dependent because of the drift of the
distribution center in the detector frame. To obtain the total PSF for
an exposure time $T_e$, this motion needs to be first inserted into
the expression of the instantaneous distribution of photons and then
integrated. When the instantaneous distribution is a
circularly symmetric Gaussian drifting at a constant speed - as is
assumed here - an analytical expression for the total PSF is
provided in \citet{Veres12}. The expression reported by these
authors (with our notations and the
correction of a sign error in their approach) 
is given in Eq.~(\ref{eqMoG-2D}) and used hereafter to represent
the total PSF (denoted $\Phi_{M^2}$) of a linearly drifting source observed with a
two-dimensional detector,
\begin{eqnarray}
\Phi_{M^2} \left( u-u_c,v-v_c\right) &=& \tilde F ~ \overline{\Phi}_{M^2}\left(
u-u_c,v-v_c\right)\nonumber\\ 
&=&\tilde F ~ \overline{\Phi}_S\left(v-v_c\right) ~
\overline{\Phi}_M\left( u-u_c\right)
\label{eqMoG-2D}
,\end{eqnarray}
where
\begin{eqnarray}
\begin{array}{r}
\overline{\Phi}_M\left( u-u_c\right) = 
\frac{P(U_2)-P(U_1)}{2L} = \frac{1}{\sqrt{\pi}L}~\int_{U_1}^{U_2}
e^{-U^2} dU
\end{array}&&\nonumber\\
\begin{array}{r}
U_1=\left(\frac{u-u_c}{\sqrt{2}\sigma}-\frac{L}{2\sqrt{2}\sigma}\right)~\text{and}~ 
U_2=\left(\frac{u-u_c}{\sqrt{2}\sigma}+\frac{L}{2\sqrt{2}\sigma}\right) 
\end{array}&&\nonumber
\end{eqnarray}
where we introduce a second coordinate system $(u,v)$ for the sensor
frame, which is right-handed orthonormal, with its $U$-axis pointing in
the source-drifting direction and its origin at the origin of the
$(x,y)$ coordinate system.  To express Eq.~(\ref{eqMoG-2D}) in the
usual $(x,y)$ coordinate system, we use the rotation between these two
coordinates systems given by Eq.~(\ref{eqc2}):
\begin{equation}
\begin{array}{rccccc}
u & = & + & x ~ \cos{\alpha} & + & y ~ \sin{\alpha}\\
v & = & -  & x ~ \sin{\alpha} & + & y ~ \cos{\alpha}
\end{array}
\label{eqc2}
,\end{equation}
where $\alpha$ is the angle measured from the $X$-axis to the
$U$-axis in the direction of the source motion.

The other parameters and functions involved in Eq.~(\ref{eqMoG-2D})
are the Gauss error function $P$ (also named probability integral),
the speed of the source $V_u$ along the $U$-axis, the distance $L$
covered by the distribution center during the whole exposure time
$T_e$ and which we call the drifting parameter or equivalently,
the elongation of the source drift ($L = V_u ~
T_e$), and the coordinates of the instantaneous PSF center $(u_c,v_c)$
in the $(u,v)$ coordinate system at the time $t=0$ corresponding to
the middle of the total exposure time $T_e$. The other parameters
($\tilde F$, $\sigma$, etc.) are similar to those of the
stationary case.

For our study, we also need an expression for the total PSF of a
linearly drifting source that is observed with a linear detector. This
expression can be easily deduced from the two-dimensional case by
assuming in Eq.~(\ref{eqMoG-2D}) that the source drifting occurs along the
$X$-axis of the linear detector and by integrating the function
$\Phi_{M^2}$ over the whole $V$-axis. Then, we obtain
Eq.~(\ref{eqMoG-1D}) as the expression of the total PSF (denoted $\Phi_M$) for a linearly
drifting source observed with a linear detector, 
\begin{eqnarray}
\Phi_{M} \left( x-x_c\right) &=& \tilde F ~ \overline{\Phi}_{M}\left(x-x_c \right)
\label{eqMoG-1D}
,\end{eqnarray}
where
\begin{eqnarray}
\begin{array}{r}
\overline{\Phi}_M\left( x-x_c\right) = 
\frac{P(X_2)-P(X_1)}{2L} = \frac{1}{\sqrt{\pi}L}\int_{X_1}^{X_2}
e^{-X^2} dX
\end{array}&&\nonumber\\
\begin{array}{r}
X_1=\left(\frac{x-x_c}{\sqrt{2}\sigma}-\frac{L}{2\sqrt{2}\sigma}\right)~\text{and}~ 
X_2=\left(\frac{x-x_c}{\sqrt{2}\sigma}+\frac{L}{2\sqrt{2}\sigma}\right) 
\end{array}&&\nonumber
\end{eqnarray}

where $x_c$ is now the coordinate of the distribution center at
instant $t=0$ (corresponding to the middle of the exposure time $T_e$)
and where the drifting parameter $L = V_x ~ T_e$ with $V_x$
the speed of the drift along the $X$-axis. We note that we can also consider the integral of $\Phi_{M^2}$ with respect to $v$ over the width of the central pixel alone, where the integrated flux is maximum. In this case, the resulting function is equal to $\Phi_{M}$ multiplied by a scale factor.

\begin{figure}[ht]
\begin{center}
\includegraphics[width=9cm]{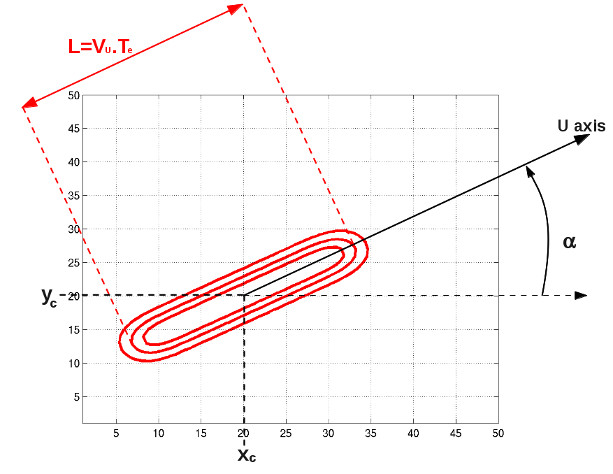}
\caption{\textit{Isophotes of the 2D MoG function characterized by
    Eq.~(\ref{eqMoG-2D}), and its main parameters for a source
    drifting at an angle $\alpha$ with respect to the X coordinate of
    the CCD. The speed of the source is $V_u$ and it
    is observed for an interval of time $T_e$. The instant $t=0$
    corresponds to the middle of the exposure time $T_e$, with
    coordinates $(x_c,y_c$). The isophotes correspond 
    to a flux level of $3/4 F_{Plateau}$, $1/2 F_{Plateau}$ and 
    $1/4 F_{Plateau}$, with the drifting parameter indicated 
    by the intermediary isophote since $L >> FWHM$.}}
\label{fig1}
\end{center}
\end{figure}

These functions $\Phi_M$ and $\Phi_{M^2}$ are what we call
drifting PSF (DPSF). We call $\Phi_M$ the
one-dimensional moving Gaussian (or 1D MoG) function, while we
name $\Phi_{M^2}$ the two-dimensional moving Gaussian (or
2D MoG function). The geometry of the 2D MoG function and its main
parameters are summarized in Fig.~\ref{fig1}.

We note that the normalized 2D MoG function ($\overline{\Phi}_{M^2}$) is the product of a normalized 1D Gaussian ($\overline{\Phi}_S$)
and a normalized 1D MoG function ($\overline{\Phi}_M$).

In this paper, $\Phi_M$ and $\Phi_{M^2}$ allow us to represent the PSF
of a moving source (such as asteroids, meteors, and artificial satellites) on a linear and a two-dimensional detector, respectively. As
extensively shown in \citet{Veres12}, the function $\Phi_{M^2}$ is one
of the most accurate PSF models for extracting astrometric data from moving sources observed with a CCD sensor. We note that for photometry a more accurate PSF model has been proposed for moving source in \citet{Fraser16}.

\section{CRLB behavior for moving sources observed with one-dimensional array detectors.}
\label{sec:StudyCR_1D}
  
\subsection{CRLB expression for the 1D MoG spread function}
\label{subsec:CR_moving_1D}  

In statistics, the CRLB gives the lower
bound for the variance of an estimated parameter: it can be represented as
the inverse of the Fisher information, which characterizes the amount
of information about this parameter contained in an observable random
variable. MSL13 established the expression of the CRLB for the astrometric
precision for a linear array detector,
with the measurement noise driven by a Poisson distribution (the adoption of this probabilistic model is common in contemporary astrometry (e.g., in Gaia, see \citet{Lindegren08})). The most generic expression for the CRLB given in MSL13 (their Eq.~(11)) can be written as follows:
\begin{equation}
\Frac{1}{\sigma_{CR}^{2}} = \sum_{i=1}^{n}\frac{\left(
    \Derpar{\tilde F_i}{x_c}\left(x_c\right)\right)^2}{\left(\tilde F_i
    \left(x_c\right)+\tilde B_{i}\right)}
\label{eqcr1}
.\end{equation}

$\sigma_{CR}^2$ is the CRLB for the variance of the PSF center $x_c$
of a source observed with a one-dimensional array. The subscript $i$
allows identifying quantities relative to the pixel of index $i$:
$\tilde F_i\left(x_c\right)$ represents the flux in pixel $i$ (in
photo-$e^-$ on the detector), whereas the background flux in the same
pixel is denoted by $\tilde B_i$ and includes contributions
from the detector such as dark-current and read-out noise ($RON$), as
well as contributions from the sky background (see
  Eqs.~(\ref{eqBG1}) and~(\ref{eqBG3})). Throughout this paper, we assume for simplicity that $\tilde B_i$ is uniform (constant)
under the source (and equal to $\tilde B$ in one pixel).

The function $\tilde{F}_i\left(x_c\right)$ involved in
Eq.~(\ref{eqcr1}) can be expressed by the integral over pixel $i$ of
the one-dimensional PSF ($\Phi$) of the source centered at $x_c$ as
follows:
\begin{equation}
\tilde{F}_i\left(x_c\right)=\int_{x_i^-}^{x_i^+}\Phi\left(x-x_c\right) dx=\tilde{F}\int_{x_i^-}^{x_i^+}\overline{\Phi}\left(x-x_c\right) dx
\label{eqcr2}
.\end{equation}

Where the integrals bounds $x_i^-$ and $x_i^+$ equal $x_i-\Delta x /
2$ and $x_i+ \Delta x /2,$ respectively (with $x_i$ the coordinate of
the center of pixel $i$ and $\Delta x$ the pixel width). The other
parameters have the same definition as in Sect.~\ref{sec:SpreadFun}:
$\tilde{F}$ is the source total flux and $\overline{\Phi}$ the normalized
PSF. We note that our function $\overline{\Phi}_S$ that we defined in the previous section is the same as the function adopted by MSL13 for the normalized PSF to express $\tilde{F}_i$ in Eq.~(\ref{eqcr1}).
 
In a similar way, we can obtain the CRLB expression for a moving
object by substituting the function $\overline{\Phi}$ in
Eq.~(\ref{eqcr2}) by the one-dimensional expression of the spread
function $\overline{\Phi}_M$ of a moving point source given by
Eq.~(\ref{eqMoG-1D}). The resulting expression for the CRLB in this case is
the following:

\begin{equation}
\sigma_{CR}^{2} = \Frac{4L^2}{\tilde
  F^2}~\Frac{1}{\sum_{i=1}^{n}\Frac{N_i^2}{D_i}}
\label{eqMoG-CR2}
,\end{equation}
where
\begin{eqnarray}
\begin{array}{rcl}
 N_i &=& \left[P\left(X_2\right)-P\left(X_1\right) \right]_{x_i^+}^{x_i^-}\\
 D_i &=& \tilde B + \frac{1}{2L} ~ \tilde F ~ I_i\\
 I_i &=& \int_{x_i^-}^{x_i^+} \left(P\left(X_2\right)-P\left(X_1\right)\right)dx\\
 &=& \frac{\sqrt{2}\sigma}{\sqrt{\pi}}\left[\left( e^{-(X_2)^2} -  e^{-(X_1)^2} \right) + \sqrt{\pi}\left(X_2 P(X_2) - X_1 P(X_1)\right)\right]_{x_i^-}^{x_i^+}
\end{array}&&\nonumber
\end{eqnarray}
and where all the parameters and functions have been defined in
Sect.~\ref{sec:SpreadFun}.

\subsection{Qualitative behavior of the CRLB expression for a linear detector.}
\label{subsec:StudyCR_CSTBGperDEG}  

In this subsection, we study the behavior of Eq.~(\ref{eqMoG-CR2}) when
the background in each pixel follows Eq.~(\ref{eqBG1}), which corresponds
to a realistic expression for the background on a CCD during an
astronomical observation,
\begin{equation}
\tilde{B}=b ~ \Delta x + D + RON^2
\label{eqBG1}
,\end{equation}

where $b$ is the sky-brightness component of the background (in units
of $e^-$/arcsec), while $D$ and $RON$ are the
dark current (mostly negligible for modern optical semi-conductor detectors) and the standard deviation of the read-out noise of the detector (in $e^-$), respectively.
We note that the requirement of a realistic estimator for the noise variance and a noise following a Poisson distribution to apply the CRLB expression~(\ref{eqMoG-CR2}) leads us to assume that the $RON$ component of the background also follows a Poisson distribution.

\begin{figure}[ht]
\begin{center}
\includegraphics[width=9cm]{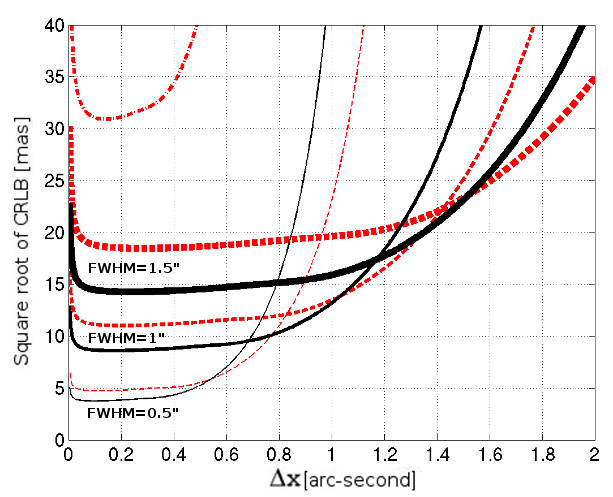}
\caption{\textit{Square root of the CRLB in mas as a function of
    detector pixel size $\Delta x$ in arcseconds when the
    background flux per pixel is given by Eq.~(\ref{eqBG1}) with
    $b=2000$ $e^-$/arcsec, $D$ is zero, and $RON=5~e^-$. 
    The solid lines (thin, normal, and bold) are for the PSF of a non-moving 
    Gaussian source with $F=6000$ $e^-$ and an image quality
with an FWHM of $0.5$, $1.0$, 
    and $1.5$", respectively (all centered on a given pixel).  
    The dashed lines correspond to the same curves computed for a slow-moving source 
    with a drifting parameter equal to the FWHM that was used to compute each line.
    The upper dash-dotted line is similar to the lower dashed line 
    (FWHM$=0.5$"), but for a source ten times fainter ($F=600$ $e^-$).}}
\label{figbcas1}
\end{center}
\end{figure}

The three solid lines (thin, normal, and bold) in Fig.~\ref{figbcas1}
show the square root of the CRLB as a function of detector pixel size for 
a non-moving source with a Gaussian PSF centered on a given pixel and 
for an FWHM of $0.5$, $1.0$ and $1.5$", respectively.  
The dashed lines correspond to the
same curves, but for a slow-moving source (with drifting
parameter $L$ equal to the FWHM) whose PSF is also centered on a given
pixel.

Figure~\ref{figbcas1} shows that for a given image quality, the 
CRLB minimum is understandably degraded by the
motion of the source. We also point out that the shapes of the dashed
curves (for a slow-moving source) and the solid curves (for a
non-moving source) are similar. In particular, we can distinguish
three regimes for the behavior of the CRLB according to the pixel size for
a slow-moving source with a constant flux $\tilde F$ and a constant
drifting parameter $L$, as described below.
\begin{itemize}
\item[$\bullet$] The oversampled phase: for pixel sizes considerably
  smaller than the image quality FWHM (in this case, $\Delta x < 0.1$ arcsec), 
  the background flux becomes
  preponderant. In this part of the curve, while the pixel size
  decreases, the part of flux that is due to the background increases
  compared to the flux of the source (since a part of the background flux is independent of the pixel size), and this leads to an increase in the CRLB
  (lower precision).
\item[$\bullet$] The undersampled phase: for pixel sizes larger than
  the image quality FWHM, the main part of the flux from the source
  is contained in very few pixels and the CRLB increases quickly. In the same way as
  for a non-moving source, the value of pixel size for
  which this degradation occurs largely depends on the centering of
  the spread function inside the pixel (see below). 
\item[$\bullet$] The well-sampled phase between the two areas described
  above ($\Delta x \sim \frac{1}{2}$ FWHM), where the CRLB reaches a
  quasi-constant floor. We note that when the contribution that
is due to
  the source of the total flux decreases, the size of the well-sampled area 
  also decreases (e.g., compare the shape of the lower dashed line with the upper 
  crossed line, which is computed for a source ten times fainter, all else being equal). 
  The reason is that as the background increases, the 
  oversampled area extends to larger pixel sizes, while the undersampled 
  area begins at an unmodified pixel size, close to the FWHM. 
  Then, for a total flux that is largely dominated by the background, the ``well-sampled floor'' 
  will collapse into a unique point corresponding to an optimal pixel size in which the
  CRLB reaches its minimum.
\end{itemize}

\begin{figure}[ht]
\begin{center}
\includegraphics[width=9cm]{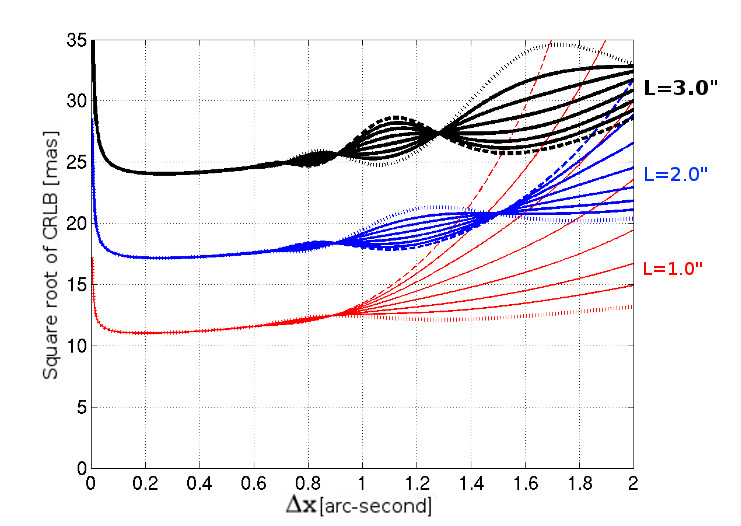}
\caption{\textit{Square root of the CRLB in mas as a function of detector pixel size $\Delta x$ in arcseconds for moving sources with $\tilde{F}=6000$ $e^-$ and where $\tilde{B}$ is given by Eq.~(\ref{eqBG1}) with $b=2000$ $e^-$/arcsec, $D$ is zero, and $RON=5$ $e^-$. The FWHM is equal to $1.0$". The three sets of curves correspond (from bottom to top) to a source drifting equal to $1.0$", $2.0$", and $3.0$", respectively. Within each set, each line corresponds to a specific value of decentering between $-0.45$ pixel (dotted lines) and $0$ pixel (dashed lines).}}
\label{figbcas3}
\end{center}
\end{figure}

We now consider the behavior of the CRLB when the PSF is not centered on a given pixel. Figure~\ref{figbcas3} shows a set of curves corresponding to the CRLB versus pixel size for an FWHM of $1.0$", for three different values of the source drift, and for different values of the PSF center inside one pixel. The lowest curves of this figure correspond - like the intermediate dashed lines in Fig.~\ref{figbcas1} - 
to the CRLB of a slow-moving source with $L$ = FWHM. 
As in the case of the stationary source (see MSL13), we see that the decentering effect on the 
CRLB of a slow-moving source is almost negligible in the oversampled and well-sampled domains,
but it plays a leading role in the undersampled case when the pixel size exceeds the FWHM. 

We now focus on the behavior of the CRLB for faster sources
(i.e., for sources with a drifting parameter $L$ larger than the FWHM).
As intuitively expected, we see in Fig.~\ref{figbcas3} that the larger
the drift of the source, the larger the degradation of the CRLB.  We also observe a number of significant oscillations of the
CRLB values according to pixel size that are not present for
stationary or slow-moving sources.  The number and amplitudes of
these oscillations increase with increasing speed of the
source. Their amplitudes form an insignificant part of the CRLB value
itself in the case of small pixel sizes (for oversampled and
well-sampled sources this effect is negligible), but the contribution becomes
important in the undersampled ``intermediate'' domain that is
bounded by pixel
sizes between the image quality FWHM and the elongation of the source
drift. In the undersampled domain that is defined by pixel sizes larger than
the drifting parameter, the effect that is due to oscillations disappears and
the decentering effect dominates the behavior of the CRLB (as for
stationary or slow-moving sources).

The oscillations of the CRLB that are observed for fast-moving sources are due to
the numerical discretization of the source PSF (by the detector array)
for which some resonances occur when the ratio between the pixel size
and the drifting parameter reaches some specific values.  Notably, we
observe some peculiar localizations (pixel size values) for which the
decentering effect is almost suppressed when the source is affected
by a specific fast drift, while this effect is preponderant when the
same source is stationary or affected by a slow drift (e.g., compare
the behavior of the CRLB for sources with drift parameters of
$1.0$" and $2.0$" at a pixel size of $1.5$" in Fig.~\ref{figbcas3}).

These peculiar locations can be of particular significance when
performing accurate astrometry of fast objects (e.g., near-Earth
objects (NEOs), artificial satellites, space debris, or meteors) even
with small telescopes whose pixel size can be substantially larger
than the typical seeing. When the speed of the
source is known, it is indeed possible (and advantageous) to adapt the
exposure time to increase or decrease the drifting parameter such that 
the decentering effect is suppressed or minimized given the pixel size 
of the camera in use.

For undersampled sources, it is not possible to simplify the expression given by Eq.~(\ref{eqMoG-CR2}) of the CRLB (the same is true for stationary sources), since no continuous approximation is practicable to avoid the large effect of the
detector-array discretization. Fortunately, the minimum value of the
CRLB is not in the undersampled domain, and the effect due to
oscillations and decentering can be neglected in the other two phases
of the curves, and especially in the oversampled domain, which is
further analyzed in the next subsection.

\subsection{CRLB approximation in the oversampled case}
\label{subsec:StudyCR_CSTBGperDEG_OScase}  

In the oversampled case, as explained in MSL13, the CRLB can be
simplified: the pixel width $\Delta x$ is considerably smaller than
the FWHM of the PSF, and then Eq.~(\ref{eqcr2}) is well approximated by
$\tilde{F}_i\left(x_c\right)=\tilde{F} ~
\overline{\Phi}\left(x_i-x_c\right)  \Delta x$; in addition, the
sum over all pixels involved in Eq.~(\ref{eqcr1}) can be approximated
by a continuous integral over the interval $]-\infty,+\infty[$. We can
    then distinguish two limiting situations: the case when the background
    flux per pixel $\tilde B$ is clearly higher than the total flux of
    the source $\tilde F$, in which case Eq.~(\ref{eqcr1}) becomes
\begin{equation}
\sigma_{CR}^{2} = \Frac{\tilde B}{\tilde F^2 ~ \Delta
  x} ~ \Frac{1}{\int_{-\infty}^{\infty}\left(\Derpar{\overline{\Phi}}{x_c}\right)^2
  dx}~~~~~~~~~\mathrm{if}~~\tilde{F}/\tilde{B}<<1
\label{eqLB}
.\end{equation}
And the case when the background flux per pixel $\tilde B$ is clearly
lower than the total flux of the source $\tilde F$, in which case
Eq.~(\ref{eqcr1}) becomes
\begin{equation}
\sigma_{CR}^{2} = \Frac{1}{\tilde
  F} ~ \Frac{1}{\int_{-\infty}^{\infty}{\frac{1}{\overline{\Phi}}
    \left(\Derpar{\overline{\Phi}}{x_c}\right)^2
    dx}}~~~~~~~~\mathrm{if}~~\tilde{F}/\tilde{B}>>1
\label{eqLF}
.\end{equation}

We note that even though these approximations have been developed by assuming
a very small pixel size, the estimates and results deduced from them
hereafter remain true in the well-sampled case with less
accuracy, of course. This is especially visible in the next figures of this
section, where the pixel size is consciously chosen to be relatively
large (close to one-third of the FWHM).

To approximate the integrals involved in Eqs.~(\ref{eqLB}) and (\ref{eqLF}) for a moving source, we first replace the normalized PSF $\overline{\Phi}$ by the 1D MoG function $\overline{\Phi}_M$ , and then we consider two distinct cases corresponding to slow- and fast-moving sources.  
 
\subsubsection{Sources with small drifting parameter}  
\label{subsubsec:StudyCR_CSTBGperDEG_OScase_SDP}  

We first study the case when the elongation of the source (as
a result of its drift) is small compared to the image quality FWHM. 
To achieve this,
we replace the reciprocals of the integrals involved in
Eqs.~(\ref{eqLB}) and (\ref{eqLF}) by their Taylor expansions in the
vicinity of a drifting parameter equal to zero (for details see
Appendix~\ref{app:MsL}). Then, for a slow-moving source,
Eqs.~(\ref{eqLB}) and (\ref{eqLF}) of the CRLB in the oversampled
regime become expression (\ref{eqLB2}) when the background dominates
the total flux, and expression (\ref{eqLF2}) when the source dominates
the total flux,
\begin{eqnarray}
\label{eqLB2}
\sigma_{CR}^{2} &=& \Frac{4\sqrt{\pi } \tilde B\sigma^3}{\tilde F ^2
  ~ \Delta x} ~ \left[1 +
  \frac{1}{2}\left(\overline{L}\right)^2 +
  \frac{1}{12}\left(\overline{L}\right)^4
+o\left[\overline{L}^{8}\right]\right]\\
&&~~~~~~~~~~~~~~~~~~~~~~~~~~~~~~~~~~~~~~~~~~~~~~~~~~~~\mathrm{if}~~\tilde{F}/\tilde{B}<<1\nonumber
\end{eqnarray}
\begin{eqnarray}
\label{eqLF2}
\sigma_{CR}^{2} &=& \Frac{\sigma^2}{\tilde F} ~ \left[ 1 +
  \frac{1}{3}\left(\overline{L}\right)^2
+o\left[\overline{L}^{8}\right]\right]~~~~~~~~~~\mathrm{if}~~\tilde{F}/\tilde{B}>>1
,\end{eqnarray}
where $\overline{L}$ is the normalized drifting parameter
equal to $\frac{L}{2\sigma}$.

The scale coefficients $\frac{4\sqrt{\pi } \tilde B\sigma^3}{\tilde F ^2 ~ \Delta x}$ and $\frac{\sigma^2}{\tilde F}$ of Eqs.~(\ref{eqLB2}) and (\ref{eqLF2}) correspond to the two expressions given by MSL13 for a non-moving source for a weak (or faint) source (MSL13 expression (39)) and for a strong (or bright) source (MSL13 expression (42)), respectively.
\begin{figure}[ht]
\begin{center}
\includegraphics[width=11cm]{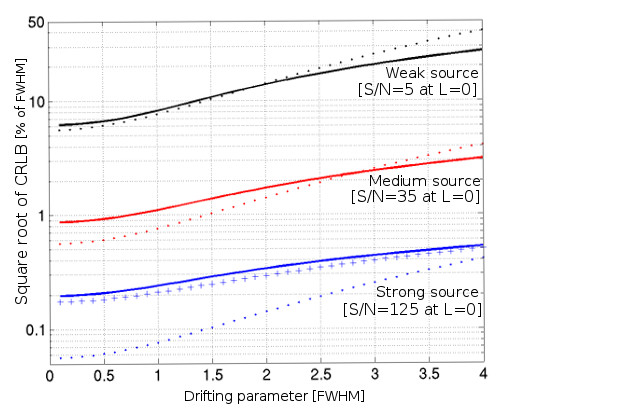}
\caption{\textit{Square root of the CRLB (in percentage of the FWHM) versus the drifting parameter of the source (in FWHM) for a detector pixel size of $0.3$" and FWHM$=1.0$" (the background is given by Eq.~(\ref{eqBG1}) with $b=2000~e^-/\text{arcsec}$, $D=0$ and RON$=5$ $e^-$). The three solid lines correspond to the exact CRLB given by Eq.~(\ref{eqMoG-CR2})  from bottom to top for a strong source ($\tilde F=60000$ $e^-$), a medium source
($\tilde F=6000$ $e^-$ ), and a weak source ($\tilde F=600$ $e^-$), respectively. The three dotted lines from bottom to top correspond to the same sources, but with the CRLB computed with the approximate expression (\ref{eqLB2}). The line with $\text{a plus sign}$ corresponds to the CRLB computed with the approximate Eq.~(\ref{eqLF2}) for the strong source. The indicated S/N is measured at peak value.}}
\label{figapp}
\end{center}
\end{figure}

For a total flux dominated by the background, we observe
for the {\it{weak source}} and the {\it{medium source}} in
Fig.~\ref{figapp} that the estimator given by Eq.~(\ref{eqLB2})
(dotted lines) provides a correct approximation of the general
behavior of the CRLB for a moving source with a drifting parameter
smaller than or equal to about twice the image quality FWHM. Of course, the
larger the background compared to the source flux, the smaller 
the difference between this estimator and the CRLB that is computed with the
exact expression (\ref{eqMoG-CR2}). For a source with a drifting
parameter larger than two, this difference increases quickly and a
better estimator is given by expression (\ref{eqLB3}) below. For a
drifting parameter equal to $\frac{3}{2}$ FWHM, the omission
of $(L^8)$-term in expression (\ref{eqLB2}) already represents
$5\%$ of the total value.

When the total flux is dominated by the source, that is, in the {\it{strong source}} case shown in Fig.~\ref{figapp}, the estimator given
by Eq.~(\ref{eqLF2}) (curve with the plus sign) provides an overall better
approximation of the behavior of the CRLB than the estimator given by
Eq.~(\ref{eqLB2}) (bottom dotted lines), as expected. This estimate
remains correct until a source with a drifting parameter equal to about
twice the image quality FWHM (for a drifting parameter equal to $\frac{3}{2 }$ FWHM, 
the neglected higher order terms $(L^8)$ and $(L^{10})$
in expression (\ref{eqLB2}) already represent $5\%$ of the total
value).

We note that at the starting point, when the speed of
the source is close to zero, the estimators given by
Eqs.~(\ref{eqLB2}) and (\ref{eqLF2}) are always lower than the exact
expression of the CRLB since the background as well as the source flux
are of course always lower than the sum of their fluxes.

Finally, we also note that for a slow-moving object that is observed in an oversampled or well-sampled regime, a global estimator of the CRLB can be taken as the maximum  of the two estimators given by Eqs.~(\ref{eqLB2}) and (\ref{eqLF2}). This global estimator is valid regardless of the ratio of $\tilde{F}/\tilde{B}$ . 
  
\subsubsection{Sources with large drifting parameters}  
\label{subsubsec:StudyCR_CSTBGperDEG_OScase_LDP}  

Second, we study the oversampled case when the drifting parameter of
the source is large compared to the image quality FWHM. Accurate
approximations of the integrals involved in Eqs.~(\ref{eqLB}) and
(\ref{eqLF}) exist in this case as well (see Appendix
~\ref{app:MbL}). Then, for fast-moving sources, the two expressions
of the CRLB in the oversampled regime become
\begin{equation}
\sigma_{CR}^{2} = \Frac{\sqrt{\pi}\tilde B\sigma}{\tilde
  F^2~\Delta x}~ L^2=\Frac{4\sqrt{\pi}\tilde B\sigma^3}{\tilde
  F^2~\Delta x}~\overline{L}^2
~~~~~~~~~~\mathrm{if}~~\tilde{F}/\tilde{B}<<1
\label{eqLB3}
\end{equation}
\begin{equation}
\sigma_{CR}^{2} = 0.55~\Frac{\sigma L}{\tilde F} =1.11 ~ \Frac{\sigma^2}{\tilde F}~ \overline{L}
~~~~~~~~~~~~~~~\mathrm{if}~~\tilde{F}/\tilde{B}>>1
\label{eqLF3}
.\end{equation}

\begin{figure}[ht]
\begin{center}
\includegraphics[width=10.5cm]{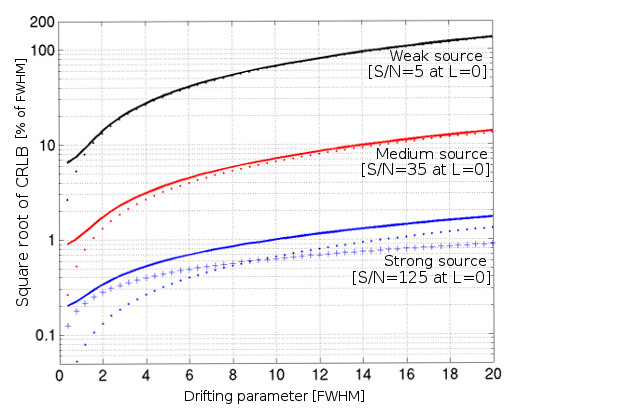}
\caption{\textit{Solid lines of this figure are similar to those of Fig.~\ref{figapp}, but for a larger drifting parameter. The lines with dots correspond to the same curves as the solid lines, but the CRLB for them has been approximated by Eq.~(\ref{eqLB3}). The line with $\text{a plus sign}$ corresponds to the CRLB approximated by Eq.~(\ref{eqLF3}) for the strongest source.}}
\label{figappBigL}
\end{center}
\end{figure}

Expressions (\ref{eqLB3}) and (\ref{eqLF3}) give good
approximations of the CRLB in the oversampled and well-sampled areas
even for a source with a relatively small drifting parameter (see
Fig.~\ref{figappBigL}). 

The approximation is less accurate for a source with an
intrinsic flux close to the background value.  This is for instance
the case of the {\it{medium source}} in Fig.~\ref{figappBigL}. For a drifting parameter of $2~$FWHM, the difference in this
source between the approximation given by expression (\ref{eqLB3}) and
$\sigma_{CR}$ given by the exact expression (\ref{eqMoG-CR2}) is
$18\%$, but this difference decreases to less than $5\%$ for a drifting
parameter of $10~$FWHM.

For the total flux dominated by the background flux, we observe that expression (\ref{eqLB3}) corresponds to twice the term in $\overline{L}^2$ of the Taylor expansion of the CRLB for a moving source with a small drifting parameter (see Eq.~(\ref{eqLB2})). We also note that this expression is equal to the estimator given by MSL13 for a non-moving source in their Eq.~(39), multiplied by the square of $\overline{L}$. In this case, $\sigma_{CR}$ evolves as $L$.

For the total flux dominated by the source, expression (\ref{eqLF3}) is equal to the estimator given by MSL13 for a non-moving source in their Eq.~(42), multiplied by $1.11\times\overline{L}$. In this case, $\sigma_{CR}$ evolves as the square root of $L$.

Finally we note that for a moving source with a drifting parameter larger than twice the image quality FWHM observed in the oversampled or well-sampled regime, a generic estimator of the CRLB is given by the maximum of the two expressions (\ref{eqLB3}) and (\ref{eqLF3}). The use of this generic estimator is particularly recommended to estimate the CRLB of sources that can be considered as bright when stationary, but that become fainter with an increase in the drifting parameter (see, for instance, the strong source
in Fig.~\ref{figappBigL}).

\subsection{Optimum exposure time for astrometry of a moving source.}
\label{subsec:OptExpTime}
For a stationary source, the relation between exposure time and
astrometric precision (according to the CRLB estimators given by the constant terms in Eqs.~(\ref{eqLB2}) and (\ref{eqLF2})) is trivial.
For a non-moving source, the longer the exposure time, the
higher the integrated flux of the source, which leads to an improved astrometric
precision (the only limit is the saturation threshold of the detector). However, for a moving source, this is no longer true. Here, the longer the exposure time, the higher the integrated flux of the source, but - and this is the main
difference to the stationary case - the larger the detector-array area covered
by the source (because the distribution
of the source flux drifts). As a result, as we increase the exposure time of a moving source, we add a comparatively larger noise that is due to the areal increase in the background.

In this subsection, we use the previously developed CRLB expressions
for a moving source to determine the optimum exposure time that allows 
reaching the best astrometric precision. We note that instead of the CRLB exact expression (\ref{eqMoG-CR2}) itself, we often use its approximations in the oversampled regime here because the CRLB minimum value is reached for pixel size values for which these simplified expressions still give a valid approximation of the CRLB (see Sect.~\ref{subsec:StudyCR_CSTBGperDEG_OScase}). As a cautionary
note, the estimate of this optimum exposure time is not correct when
the pixel size falls into the intermediate or undersampled
regimes. 

\begin{figure}[t]
\begin{center}
\includegraphics[width=11cm]{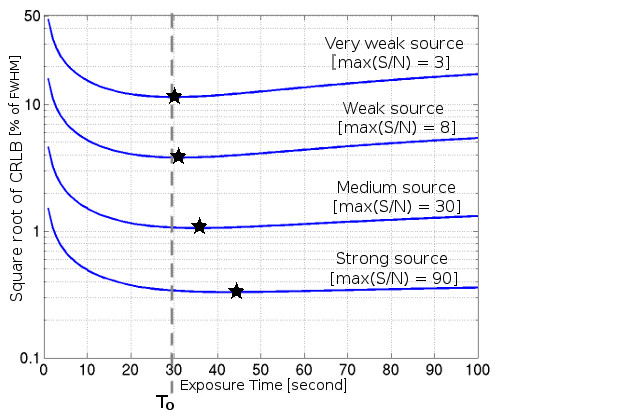}
\caption{\textit{Square root of the CRLB versus exposure time 
    for sources with a speed of $2$"/min and for which the fluxes per minute $f_s$ 
    from top to bottom of the figure are  $600$, $2000$, $10000,$ and 
    $60000$ $e^-$/min (FWHM$=1.0$", $\Delta x=0.3$", 
    $b_s=2000$ $e^-$/arcsec/min, $d=0,$ and RON$=0$), respectively. 
    The star symbols correspond to the optimum exposure time for each source. 
    The vertical dashed line is the lower limit $T_O$ of the optimum exposure 
    time when the flux is completely dominated by the sky background.}}
\label{figbcasT2}
\end{center}
\end{figure}

In this scenario, the background is time dependent, and
its expression becomes
\begin{equation}
\tilde{B}=b_1~T_{e} + b_0 = \left(b_s ~ \Delta x + d\right) T_{e}+RON^2
\label{eqBG2}
,\end{equation}

where $b_1 ~ T_e$ is the background part depending on exposure time $T_e$ , while $b_0$ is the time-independent part. $b_1$ depends on $b_s$, which is the
sky component (in units of $e^-$/arcsec/sec) and $d,$ which is the
dark-current component (in units of $e^-$/sec). $b_0$ depends on $RON,$ which is assumed
to be independent of exposure time. We recall that the relation between the source total flux and exposure time is $\tilde{F}=f_s ~ T_{e}$ , while the relation between source speed and its drifting parameter is given by $L=V_x~ T_{e}$.  
With these new notations, we compute the CRLB as a function of exposure time for four 
sources with a peak S/N bewteen $3$ and $90$ and a similar speed 
of $2.0$"/min. The pixel size, image FWHM, and background flux (whose RON component is negligible) are kept constant (see Fig.~\ref{figbcasT2}).

First we observe that the behavior of all CRLB curves is similar: with the increase in exposure time, the CRLB first decreases, then reaches a quasi-constant floor, and finally, unlike in the case of stationary sources, it reaches a turning point (where its minimum value is attained) and then starts to increase. 
When the RON is negligible (as for Fig.~\ref{figbcasT2}), the turning point is at an exposure time close to $30$ seconds, which corresponds to an elongation of the source drift close to the FWHM (equal to $1.0$" in the figure). When the RON component of the background increases, the turning point is located at a slightly longer exposure time.   

We also note that the brighter the source, the longer the exposure time that is required to reach the minimum value of the CRLB (see the strong source case in Fig.~\ref{figbcasT2}).  For the theoretical limiting case when the background flux is zero, this exposure time is infinity. An estimate of the corresponding minimum value of the CRLB in this theoretical case can be given accurately by expression (\ref{eqLCRTe}), which is the limit when the exposure time approaches infinity in Eq.~(\ref{eqLF3}), which is valid for
a high value of the drifting parameter,\begin{equation}
\label{eqLCRTe}
\lim\limits_{T_e \rightarrow +\infty} \sigma_{CR}^{2}
= 0.55 ~ \Frac{\sigma}{f_s} ~ V_x = 0.24 ~ \text{FWHM} ~ \Frac{V_x}{f_s} 
.\end{equation}

Expression (\ref{eqLCRTe}) is of particular significance. It gives a
simple way to estimate an absolute limit for the astrometric
precision of all moving sources with the knowledge of only three
parameters, which are the FWHM, the flux of the source
in electrons per second, and the source speed in arcseconds per
second.

By contrast, the fainter the source, the smaller the size of the CRLB floor, and the more the optimum exposure time converges toward a lower limit that is close to the exposure time that corresponds to a drifting parameter that is equal to the image quality FWHM. 
We call $T_o$ the lower limit of the optimum exposure time (see Fig.~\ref{figbcasT2}) and $L_o$ the corresponding drifting parameter $L_o = V_x ~  T_o$. 
We can calculate $T_o$ as a power series by considering the roots of the derivative of expression~(\ref{eqLB}) with respect to the exposure time when $\overline{\Phi}$ is replaced by $\overline{\Phi}_M$. Expression (\ref{eqLB}) in the oversampled area yields the limit of the expression (\ref{eqMoG-CR2}) of the CRLB when the background dominates the total flux, and Fig.~\ref{figbcasT2} shows that it is precisely in this regime that a lower limit of the optimum exposure time is achieved. Then, we can express $T_o$ as follows (see Appendix~\ref{app:AT_o} for more details):
\begin{eqnarray}
\label{eqTo}
&&T_o = \left[ 0.95 + 0.66\mu_b - 1.12\mu_b^2 + 2.75\mu_b^3\right] ~ T_s
,\end{eqnarray}
where $T_s$ is the exposure time corresponding to a drifting parameter that is equal to the image quality FWHM  ($T_s=\text{FWHM}/V_x$), and $\mu_b$ measures the ratio between the time-independent and the time-dependent parts of the background flux for an exposure time equal to $T_s$ ($\mu_b=b_0/(b_1\cdot T_s)$). 
The numerical coefficients in Eq.~(\ref{eqTo}) have been confirmed by an estimate of $T_o$ based on a binary search algorithm applied on the exact expression (\ref{eqMoG-CR2}): for $\mu_b$ below $0.2$, the agreement is better than one percent, but for $\mu_b$ larger than $0.4,$ this expression of $T_o$ is no longer valid. In particular, this means that when the RON component is an important part of the background, the lower limit $T_o$ of the optimum exposure time cannot be given by Eq.~(\ref{eqTo}) and has to be determined numerically from the Eq.~(\ref{eqMoG-CR2}).

To obtain an estimate of the CRLB value for an exposure time equal to
$T_o$, we replace the parameter $\overline{L}$ by $T_oV_x/(2\sigma)$ in
expressions (\ref{eqLB2}) and (\ref{eqLF2}) for the case when the total
flux is dominated by the background, and the case when the total flux
is dominated by the source, respectively (these CRLB estimators can be
used here since we have shown that they give correct results even for a
drifting parameter $\overline{L}$ equal to one and a half times
the FWHM). We obtain the following two expressions:
\begin{eqnarray}
\label{eqLBo}
\sigma_{o}^{2} &=& \Frac{4\sqrt{\pi } \tilde B\sigma^3}{\tilde F ^2
  ~ \Delta x} \left[ 1.76 + 1.24\mu_b - 1.41\mu_b^2 \right] ~~~ \mathrm{if}~\tilde{F}/\tilde{B}<<1\\
\label{eqLFo}
\sigma_{o}^{2} &=& \Frac{\sigma^2}{\tilde F} \left[ 1.42 + 0.58\mu_b - 0.78\mu_b^2 \right]\nonumber\\
&=& \text{FWHM}\Frac{V_x}{f_s}\left[0.27 - 0.08\mu_b + 0.22\mu_b^2 \right]~~\mathrm{if}~\tilde{F}/\tilde{B}>>1
.\end{eqnarray}
When $T_o$ is used as exposure time and when the RON is a negligible part of the total background, we can deduce from Eqs.~(\ref{eqLBo}) and (\ref{eqLFo}) that the degradation of $\sigma_{CR}$ for a moving source is between $19\%$ and $33\%$ of the $\sigma_{CR}$ of the same source with no motion observed during the same exposure time. When the RON is not a negligible part of the total background ($\mu_b$ different from zero), the CRLB degradation that is due to the motion of the source is slightly larger (since $T_o$ is longer).  

When the total flux is dominated by the source (as for the strong
sources shown in Fig.~\ref{figbcasT2}), the CRLB of a moving source
can reach a value slightly below the value given by expression
(\ref{eqLFo}). From the comparison between this expression and
Eq.~(\ref{eqLCRTe}), however, we deduce that the maximum possible improvement
for $\sigma_o$ is $7\%$, and only for a theoretical measurement
without background, and with an exposure time equal to infinity.

\begin{figure}[ht]
\begin{center}
\includegraphics[width=9cm]{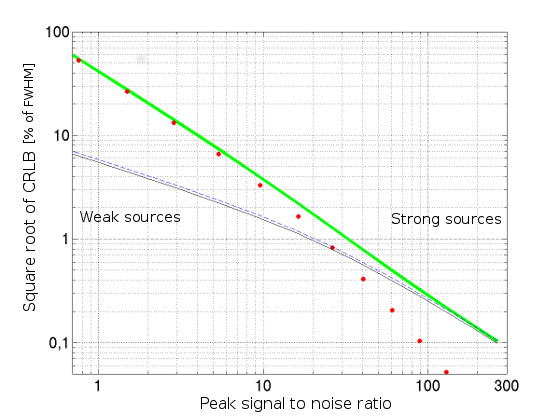}
\caption{\textit{Minimum CRLB versus the peak S/N for a moving source ($V_x=2$"/min, $f_s$ between $30$ and $500000$ $e^-$/min, FWHM$=1.0$", $\Delta x=0.3$", and $b_s=2000$ $e^-$/arcsec/min). The CRLB expressions used are the exact Eq.~(\ref{eqMoG-CR2}) (bold line), Eq.~(\ref{eqLBo}) (dots), Eq.~(\ref{eqLFo}) (dashed line), and the limiting expression (\ref{eqLCRTe}) (thin solid line).}}
\label{figbcasT3}
\end{center}
\end{figure}

In order to check the validity range of the previous expressions, we compute with the help of a binary search algorithm applied to Eq.~(\ref{eqMoG-CR2}) the exact minimum CRLB for moving sources with a speed of $2$"/min and with fluxes varying between $f_s=30$ $e^-$/min and $f_s=500000$ $e^-$/min. 
These sources are observed with a CCD pixel size of $0.3$", an image quality FWHM of
$1.0$", and with a sky-background component $b_s=2000$ $e^-$/arcsec/min (the dark and the RON background components are equal to zero). Depending on the source 
brightness, the peak S/N varies between $0.7$ and $250$. Then, we plot in Fig.~\ref{figbcasT3} for each S/N value the exact minimum CRLB value obtained by this numerical method (bold line) and their estimates with expression (\ref{eqLBo}) (dots), with expression (\ref{eqLFo}) (thin dashed line), and with expression (\ref{eqLCRTe}) (thin solid line).

For weak sources (left part of Fig.~\ref{figbcasT3}) the exact determination and the estimate made with expression (\ref{eqLBo})
agree well. When the source is stronger (right part of the figure), the agreement is better with the estimate deduced from expression (\ref{eqLFo}), and except for the strongest source of this example ($S/N=250$), the value provided by this expression is below the exact minimum CRLB value. For the strongest source the estimator given by Eq.~(\ref{eqLFo}) is just above the exact value (less than $5\%$ higher). The unreachable limit given by expression (\ref{eqLCRTe}) is always below the exact value and all the other estimates, as expected.\\

\underline{Optimum exposure time for a set of observations}

Most of the time, the observation of a moving source consists of $N$
consecutive images of the source taken with a similar exposure time
$T_e$. If during $T_a$ (the duration of the whole data set) the
meteorology, the instrumental conditions, and the linear motion of the
target remain unchanged, then the lower bound of the final 
variance of the astrometric precision of the whole data set is given by $\sigma_i^2/N$ (where $\sigma_i^2$ is the variance of the astrometric precision of one
independent image). Note that it is beyond the scope of this paper to develop a method to combine the single measurements. Hereafter we only assume that an ideal method exists and deduce a lower bound for the astrometric precision of the "mean measurement".

The CRLB formalism can help us to estimate the optimum exposure time of a single image,
which maximizes the astrometric precision of a whole data set. Let
$T_N$ be this optimum exposure time, the number of frames $N(T_N)$
contained in the data set are $(T_a+\Delta_T)/(T_N+\Delta_T)$,
where $\Delta_T$ is the time delay between consecutive
exposures. Then, the CRLB for the whole data set $\sigma_a^2(T_N)$
equals $\sigma_i^2(T_N)/N(T_N)$ (where $\sigma_i^2(T_N)$ is now the
CRLB of one image).

In what follows we develop a method to graphically locate the
optimum exposure time $T_N$ for a given source, when its characteristics ($V_x$ and $f_s$) as well as the observational conditions (FWHM, $T_a$, $\Delta x$, $\Delta_T$, $b_s$, $d,$ and RON) are assumed to be known.  We note that finding the optimum exposure time $T_N$ that minimizes $\sigma_a$ is equivalent to finding the value of $T_N$ that maximizes the ratio of $\sigma_a^2(T_s)$ to $\sigma_a^2(T_N)$, where $T_s$, defined previously, is the exposure time corresponding to a drifting elongation equal to the FWHM. With the above notation, we can write this ratio as follows:
\begin{eqnarray}
\label{eqRatioCR}
\mu_{CR}^2(T_N)=\Frac{\sigma_a^2(T_s)}{\sigma_a^2(T_N)}=\Frac{\sigma_i^2(T_s)}{N(T_s)}\Frac{N(T_N)}{\sigma_i^2(T_N)}=\Frac{\sigma_i^2(T_s)}{\sigma_i^2(T_N)}\Frac{T_s+\Delta_T}{T_N+\Delta_T}&&
.\end{eqnarray}
We note that the choice of the exposure time $T_s$ for the numerator of this ratio is arbitrary and that the same method can be applied with any fixed exposure time.
The advantage of considering this ratio instead of $\sigma_a$ itself
is that this ratio is independent of the total duration $T_a$ of the
data set. In return, with this method, $N(T_N)$ is not an integer
a priori, and $T_N$ will need to be slightly increased or decreased to
reach the closest integer $N$ that minimizes $\sigma_a$.

\begin{figure}[ht]
\begin{center}
\includegraphics[width=9cm]{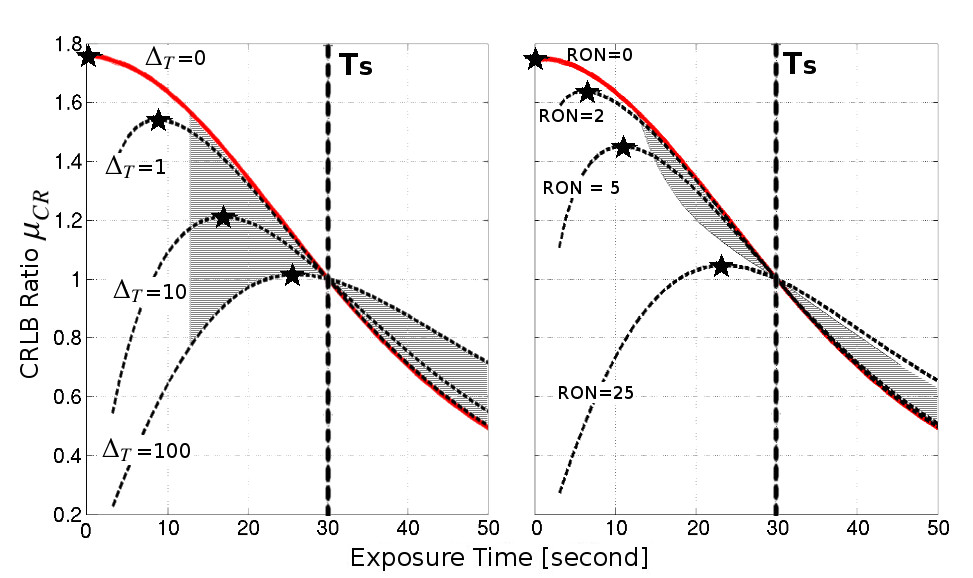}
\caption{\textit{CRLB ratio versus exposure time for a moving source
    with $V_x=2$"/min, $b_s=2000$ $e^-$/arcsec/min, FWHM$=1.0$", $\Delta
    x=0.3$", $d=0$, and $f_s=800$ $e^-$/min. The star symbols
    correspond to the maximum of each curve. The vertical dashed line
    is $T_s$, the exposure time corresponding to a drifting parameter
    equal to the FWHM (left: RON equals zero and $\Delta_T$ equals $0$,
    $1$, $10,$ and $100$ seconds; right: $\Delta_T$ equals zero and RON
    equals $0$, $2$, $5,$ and $25$ $e^-$/pix). The hatched areas
    correspond to exposure times long enough to allow source
    detection (here, the source is assumed to be detectable for a peak 
    S/N higher than $3.5$).}}
\label{figbcasT4}
\end{center}
\end{figure}

We compute and plot in Fig.~\ref{figbcasT4} the CRLB ratio of
Eq.~(\ref{eqRatioCR}) (where $\sigma_i$ is replaced by its exact
expression (\ref{eqMoG-CR2})) according to the exposure time for a
moving source with a drifting speed of $2$"/min observed with an FWHM
of $1.0$". The vertical lines at $T_e=30$ seconds correspond to $T_s$
(which is close to $T_o$ , but independent of RON, in contrast to $T_o$). 
In the left panel of the figure, we assume
that RON is zero while the time delay $\Delta_T$ varies, and
conversely, in the right panel of the figure, we assume that
$\Delta_T$ is zero while RON varies. The star symbols correspond to
the maximum of the CRLB ratio, and the abscissa gives the values of
$T_N$ corresponding to each condition.

We observe that the maximum of the limiting curve when RON and
$\Delta_T$ tend to zero (the two identical bold curves in the left and
right panels) is reached for an exposure time equal to zero, which
makes no sense. Another criterion has indeed to be taken into
account: the optimum exposure time (according to the CRLB estimator)
has to be long enough to allow the detection of the
source. As a trivial example, we consider that a source
is detected if the peak S/N is higher than a
fixed threshold characterizing the reduction pipeline capacity (fixed to $3.5$ hereafter). 

For instance, with the parameters of Fig.~\ref{figbcasT4}, the
minimum exposure time needed to reach an S/N higher than
 $3.5$ is $13$ seconds when the RON is negligible (left panel), and it
 increases when the RON increases (for a RON larger than $21$
 $e^-/pix$, the S/N is always below the threshold).
 Then, by adding this information concerning the detectability of the
 source in Fig.~\ref{figbcasT4} (hatched areas), we can easily
 determine for each observational constraint a ``realistic''
 exposure time $T_N$ for the whole data set by restricting the search
 of the CRLB ratio maximum to these hatched areas.

We see that $T_N$ increases quickly with the rise of $\Delta_T$ (left
panel) or with the rise of RON (right panel): with $\Delta_T$ equal
to only one-thirtieth of $T_s$, $T_N$ already equals one-third of
$T_s$. We also observe that this increase seems bounded by $T_s$. 
Indeed, $T_N$ is logically shorter than the optimum exposure time of a single image (minimizing $\sigma_i$), and we have shown that even though this exposure time can be longer than $T_o$ (or $T_s$), the value of $\sigma_i^2(T_o)$ is already close to the minimum of $\sigma_i$.    
We note that this bound does not exist for a stationary sources.

We do not show a power series for $T_N$ (as done for $T_o$ in
Eq.~(\ref{eqTo})) because the location of the maximum of
Eq.~(\ref{eqRatioCR}) is the result of a complex interaction between
several parameters.  However, the simplest and more straightforward
numerical approach - an estimate of $T_N$ that is based on a binary search
algorithm applied to the search of the maximum of the CRLB ratio -
allows us to compute $T_N$ and then to deduce the number of images $N$
that will optimize the CRLB of a whole set of observation of duration
$T_a$. This is particularly important when planning astrometric observations 
of moving targets. We note that This method can also be used without any modification
to compute $T_N$ when the $N$ images are stacked before the
reduction process (but the minimum exposure time for allowing
detections is shorter since the S/N is multiplied
by $\sqrt{N}$).

\section{CRLB behavior for sources observed with a two-dimensional sensor}
\label{sec:CR_moving_2D}  
This part extends the results of the previous section to two-dimensional detectors in three steps: extension of the generic CRLB expressions for these detectors, application of these expressions to a stationary source, and finally, in order to reach the main objective of our study, application of these expressions to the case of a source with linear and constant motion.

\subsection{Generic CRLB expressions for astrometric precision with two-dimensional detector arrays}
\label{subsec:CR_GEN_2D} 
A two-dimensional array detector of $n_x$ columns and $n_y$ rows can
be considered as a linear-array detector of $n_x\times n_y$ pixels. Then, by
keeping the same formalism as in the previous section, we can
deduce from Eq.~(\ref{eqcr1}) the expression below which
this corresponds to the CRLB for the astrometric precision along the
$X$-axis in the case of a two-dimensional detector array:
\begin{equation}
\frac{1}{\sigma_{CR_X}^{2}} = \sum_{j=1}^{n_y}\sum_{i=1}^{n_x}\frac{\left(
    \Derpar{\tilde F_{i,j}}{x_c}\left(x_c,y_c\right)\right)^2}
    {\left(\tilde F_{i,j}\left(x_c,y_c\right)+\tilde B_{i,j}\right)}
\label{eqcr2D-1}
.\end{equation}
$\sigma_{CR_X}^2$ is the CRLB for the variance of the coordinate $x_c$
of the PSF center for a source observed with a two-dimensional
array. The subscripts $i$ and $j$  now allow us to identify quantities
relative to the pixel in column $i$ and row $j$. The function
$\tilde{F}_{i,j}\left(x_c,y_c\right)$ represents the flux contained in
the pixel that is located in column $i$ and row $j$, and can be calculated
through the integral of the normalized two-dimensional PSF
($\overline{\Phi}$) of the source centered at coordinates
$(x_c,y_c)$. The expression for $\tilde{F}_{i,j}$ is
\begin{equation}
\tilde{F}_{i,j}\left(x_c,y_c\right)=\tilde{F}\int_{y_j^-}^{y_j^+} \int_{x_i^-}^{x_i^+}
\overline{\Phi}\left(x-x_c,y-y_c\right) dx dy
\label{eqcr2D-2}
,\end{equation}
where $x_i^-=x_i-\Delta x/2$, $x_i^+=x_i+\Delta x/2$, $y_j^-=y_j-\Delta y/2$ and $y_j^+=y_j+\Delta y/2$ (with $\Delta x$ and $\Delta y$ the pixel width along $X$ and $Y,$ respectively). The background flux in the same pixel is called $\tilde{B}_{i,j}$. As previously, the background is uniform under the source (equal to $\tilde{B}$ in one pixel) and can be divided into three components: the $RON$ and the dark current ($D$), which are independent of the pixel size, and the sky-brightness component $b,$ which was proportional to the pixel size in the one-dimensional case (Eq.~(\ref{eqBG1})) and which is proportional to the pixel area now in the two-dimensional case (Eq~(\ref{eqBG3})),
\begin{equation}
\tilde{B}=b~ {\Delta x}^2 + D + RON^2
\label{eqBG3}
.\end{equation}

\underline{CRLB approximations for the oversampled case}

Similar to the case of the one-dimensional detector-array, when the
pixel sizes in X and Y are small compared to the image quality FWHM,
the flux in one pixel given by expression~(\ref{eqcr2D-2}) is well
approximated by
$\tilde{F}_{i,j}=\tilde{F}~\overline{\Phi}\left(x_i-x_c,y_j-y_c\right)\Delta
x \Delta y$ and the sums over all pixels present in CRLB
expression~(\ref{eqcr2D-1}) can be replaced by continuous integrals
over the whole space. We can then distinguish two extreme scenarios:
the case when the background flux per pixel $\tilde B$ is clearly
higher than the total flux of the source $\tilde F$ (faint sources),
in which case Eq.~(\ref{eqcr2D-1}) becomes
\begin{equation}
\sigma_{CR_X}^{2} = \Frac{\tilde B}{\tilde F^2~\Delta x \Delta
  y}~
\Frac{1}{\iint_{-\infty}^{\infty}\left(\Derpar{\overline{\Phi}}{x_c}\right)^2
  dx dy}~~~~~~~~~\mathrm{if}~~\tilde{F}/\tilde{B}<<1
\label{eqLB-2D}
.\end{equation}
And the case when the background flux per pixel $\tilde B$ is clearly
lower than the total flux of the source $\tilde F$ (bright sources),
when Eq.~(\ref{eqcr2D-1}) becomes\begin{equation}
\sigma_{CR_X}^{2} = \Frac{1}{\tilde
  F}~\Frac{1}{\iint_{-\infty}^{\infty}{\frac{1}{\overline{\Phi}}
    \left(\Derpar{\overline{\Phi}}{x_c}\right)^2 dx
    dy}}~~~~~~~~\mathrm{if}~~\tilde{F}/\tilde{B}>>1
\label{eqLF-2D}
.\end{equation}

More generally, hereafter, we denote $\sigma_{CR_U}$ as the CRLB for
the astrometric precision along any direction $U$ in the reference
plane of the CCD. The expression for $\sigma_{CR_U}$ would be given by
Eqs.~(\ref{eqcr2D-1}), (\ref{eqLB-2D}), or (\ref{eqLF-2D}) (depending
on the case involved), with the derivative of $\tilde{F}_{i,j}$ (or
$\overline{\Phi}$) with respect to $x_c$ replaced by the derivative
with respect to $u_c$ (where $u_c$ is the component of the
instantaneous PSF center along the $U$-axis). For
expressions (\ref{eqLB-2D}) and (\ref{eqLF-2D}), the integral over the
whole space can also be performed with any coordinate system $(u,v)$
instead of the $(x,y)$ system adopted here since the pixel sizes are
smaller than the FWHM.

\subsection{Stationary sources on a two-dimensional array}
\label{subsec:CR_stat_2D}
The expression of the normalized PSF $\overline{\Phi}_{S^2}$ of a stationary source that is observed with a 
two-dimensional detector array is given by Eq.~(\ref{eqS-2D}) of Sect.~\ref{sec:SpreadFun}. Then, by using this normalized PSF in Eq.~(\ref{eqcr2D-2}) to compute $\tilde{F}_{i,j}$ and its derivative and by substituting them into Eq.~(\ref{eqcr2D-1}), we obtain the following expression for
the CRLB along the $X$-axis for a stationary source that is observed with a two-dimensional detector array:
\begin{equation}
\sigma_{CR_{X}}^2=\Frac{4\pi^2\sigma^4}{\tilde{F}^2}~\Frac{1}{\sum_{j=1}^{n_y}
    \sum_{i=1}^{n_x}\Frac{(F^j ~ N_i)^2}{D_i^j}}
\label{eq-stat-CR2D}
,\end{equation}
where
\begin{eqnarray}
\begin{array}{rcl}
F^j &=& \int_{y_j^-}^{y_j^+} e ^{-\frac{1}{2}\left(
  \frac{y-y_c}{\sigma}\right)^2} dy =\frac{\sqrt{2\pi}\sigma}{2}\left[
  P\left(\frac{y_j^+-y_c}{\sqrt{2}\sigma}\right) -
  P\left(\frac{y_j^--y_c}{\sqrt{2}\sigma}\right) \right] \\
N_i &=& \left[e ^{-\frac{1}{2}\left(
    \frac{x-x_c}{\sigma}\right)^2}\right]_{x_i^+}^{x_i^-}\\
D_i^j &=& \tilde B + \Frac{\tilde{F}}{2\pi\sigma^2} F^j ~ I_i\\
I_i &=&\int_{x_i^-}^{x_i^+} e ^{-\frac{1}{2}\left(
  \frac{x-x_c}{\sigma}\right)^2} dx =\frac{\sqrt{2\pi}\sigma}{2}\left[
  P\left(\frac{x_i^+-x_c}{\sqrt{2}\sigma}\right) -
  P\left(\frac{x_i^--x_c}{\sqrt{2}\sigma}\right) \right.]
\end{array}\nonumber 
\end{eqnarray}
Hereafter, we assume that the pixel sizes along $X$ and $Y$ axes are equal
($\Delta x=\Delta y$).

The three solid lines in Fig.~\ref{fig2DF} (thin, normal, and bold) computed with Eq.~(\ref{eq-stat-CR2D}) show the square root of the CRLB as a function of detector pixel size for a non-moving source centered on a given pixel and observed with a two-dimensional detector for an FWHM of $0.5$", $1.0$", and $1.5$", respectively.
We note that the overall behavior of these three lines is very similar to the three solid curves plotted in Fig.~\ref{figbcas1} for a non-moving source that is observed with a one-dimensional detector and based on Eq.~(21) of MSL13. In particular, the three distinct regimes (oversampled, well-sampled, and undersampled) also exist in the two-dimensional case. We can observe, however, that in the two-dimensional case the oversampled regime is slightly more extended, while the well-sampled area begins for pixel sizes
that are slightly larger.

\underline{Oversampled case approximations:}

The CRLB expressions (\ref{eqLB-2D}) and (\ref{eqLF-2D}) with
$\overline{\Phi}_{S^2}$ as normalized PSF allow us to produce simplified expressions for the CRLB of a stationary source that
is observed with a two-dimensional array detector (the values of the reciprocals of the integrals involved in Eqs~(\ref{eqLB-2D}) and (\ref{eqLF-2D}) for $\overline{\Phi}_{S^2}$ are given in Appendix~\ref{app:S} by the functions $I_{S_1^2}$ and $I_{S_2^2}$ , respectively). 
Thus, the expression (\ref{eq-stat-CR2D}) is well approximated by expression (\ref{eqLB2_2DG}) when the background dominates the total flux, and by expression
(\ref{eqLF2_2DG}) when the source dominates the total flux,
\begin{eqnarray}
\label{eqLB2_2DG}
\sigma_{CR_{X}}^{2} &=& \frac{2\sqrt{\pi}\sigma}{\Delta x}~
\sigma_{CR_{1D}^{B}}^{2}=\Frac{8 \pi \tilde B\sigma^4}{\tilde F
  ^2~{\Delta x}^2} ~~~~~~\mathrm{if}~~\tilde{F}/\tilde{B}<<1
\end{eqnarray}
\begin{eqnarray}
\label{eqLF2_2DG}
\sigma_{CR_{X}}^{2} &=& \sigma_{CR_{1D}^{F}}^{2} =
\Frac{\sigma^2}{\tilde F} ~~~~~~\mathrm{if}~~\tilde{F}/\tilde{B}>>1
,\end{eqnarray}
where $\sigma_{CR_{1D}^B}$ and $\sigma_{CR_{1D}^F}$ correspond to the CRLB expressions given by MSL13 in their Eqs.~(39) and (42), respectively, for a stationary source that is observed with a one-dimensional detector. When the background dominates the total flux, we observe that the CRLB in the two-dimensional case is equal to the CRLB
in the one-dimensional case multiplied by a factor that is inversely
proportional to the pixel size and larger than 1 when $\Delta x <$FWHM. This factor explains that the degradation of CRLB in an oversampled regime occurs for a larger pixel size in the two-dimensional case than in the one-dimensional case. When the source dominates the total flux, the CRLB expression remains unchanged.

We also note that even though the above expressions (\ref{eq-stat-CR2D}), (\ref{eqLB2_2DG}), and (\ref{eqLF2_2DG}) are obtained for the CRLB along the $X$-axis, they remain rigorous and unchanged for the CRLB along any direction, since the PSF has been
assumed to be circularly symmetric.
 
\subsection{CRLB of a moving source observed with a two-dimensional digital-detector array}
\label{subsec:CR_mov_2D} 

Finally, we present in this subsection the CRLB behavior for a moving source observed with a two-dimensional detector array.  
As a PSF of the moving source, we use $\overline{\Phi}_{M^2}$, the 2D MoG function given by Eq.~(\ref{eqMoG-2D}) 
and presented in Sect.~\ref{sec:SpreadFun}.  This function is not circularly symmetric, and therefore the CRLB is not the same along all directions. However, compared to the stationary case, the main modification of the CRLB behavior is, of course, in the drifting direction. Thus we first study the CRLB along the drifting direction and show the similarities and differences with the results obtained in Sect.~\ref{sec:StudyCR_1D} for a linear detector. Then, we study the CRLB along the direction normal to the motion of the object - which is of course not present in the linear case.

\subsubsection{CRLB along the drifting direction}
\label{subsubsec:CR_mov_2D_U} 
Following the conclusions of Sect.~\ref{subsec:CR_GEN_2D}, the generic
expression for the CRLB along the drifting direction (i.e., along the
$U$-axis) is given by expression (\ref{eqcr2D-1}), where the
normalized PSF $\overline{\Phi}$ is replaced by
$\overline{\Phi}_{M^2}$ and where the partial derivative of
$\overline{\Phi}$ with respect to $x_c$ is replaced by the derivative
of $\overline{\Phi}_{M^2}$ with respect to $u_c$. The resulting
expression for the CRLB along the drifting direction is the following:
\begin{equation}
\sigma_{CR_U}^{2} =\Frac{4\pi^2\sigma^4
  L^2}{\tilde{F}^2}~\Frac{1}{\sum_{j=1}^{n_y}\sum_{i=1}^{n_x}\frac{\left(N_i^j\right)^2}{D_i^j}}
\label{eq-mov-CR2D}
,\end{equation}
where now
\begin{eqnarray}
\begin{array}{rcl}
N_i^j &=& \int_{y_j^-}^{y_j^+} \int_{x_i^-}^{x_i^+} e
^{-\frac{1}{2}\left( \frac{v-v_c}{\sigma}\right)^2} \left(
e^{-U_2^2}-e^{-U_1^2}\right) dx dy\\
D_i^j &=& \tilde B + \frac{\tilde{F}}{2\sqrt{2\pi}\sigma L} I_i^j\\
I_i ^j&=&\int_{y_j^-}^{y_j^+} \int_{x_i^-}^{x_i^+} e
^{-\frac{1}{2}\left(
  \frac{v-v_c}{\sigma}\right)^2}\left(P(U_2)-P(U_1)\right) dx dy.
\end{array}\nonumber 
\end{eqnarray}

In contrast to the previous CRLB expressions, the explicit integrals cannot be easily avoided
here because the
integrals have to be performed in the pixels frame
(i.e., $(x,y)$-coordinate system), while the 2D MoG function is only
multiplicatively separable in the $(u,v)$-coordinate system. It
follows from this that the computing time for this CRLB expression is
longer than for the previous one.

Equation~(\ref{eq-mov-CR2D}) can be seen as an extension of the CRLB expression (\ref{eqMoG-CR2}) for a two-dimensional array detector, and the next figure allows us to compare the behavior of these two CRLB expressions.
\begin{figure}[ht]
\begin{center}
\includegraphics[width=9cm]{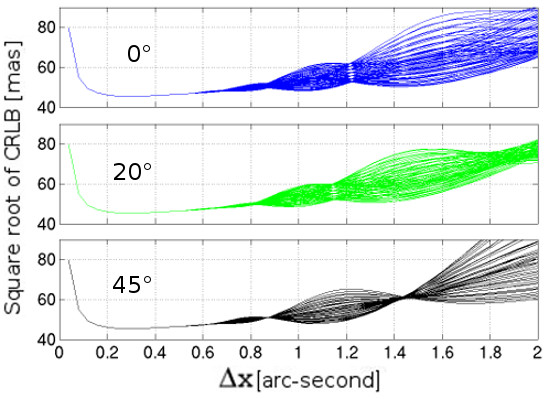}
\caption{\textit{Square root of the CRLB in mas for a moving source
    observed with a two-dimensional array detector as a function of
    detector pixel size $\Delta x$ in arcseconds. The figure is
    divided into three sub-figures, each one corresponding to a specific
    orientation of the source drift with respect of the detector-array
    frame. Within each sub-figure, each line corresponds to two
    different values of decentering along X and Y axis between $-0.46$ and $0$ pixel. 
    The parameter values are $L=3.0$", FWHM$=1.0$", $\tilde{F}= 6000$~$e^-$,
    $D=0$~$e^-$/pix, $RON=5$~$e^-$/pix, and $b=6000$~$e^-/$arcsec$^2$.}}
\label{figmov_2D}
\end{center}
\end{figure}

Figure~\ref{figmov_2D} shows three sets of curves corresponding to the
CRLB computed with Eq.~(\ref{eq-mov-CR2D}) versus pixel size for an
FWHM of $1.0$" and drifting parameter $L$ of $3.0$" for different
values of the PSF offset from the pixel center and for three different
values of the inclination angle ($\alpha$, see Fig. \ref{fig1}) of the
DPSF.

We first compare the curves for $\alpha=0^o$ in the top row of
Fig.~\ref{figmov_2D} (DPSF drift aligned with the detector $X$-axis) with the top curves ($L=3.0$") of Fig.~\ref{figbcas3}. For these two graphs, the CRLB is computed for the same source as was observed in the same conditions, but with a two-dimensional detector array in the first case (Eq.~(\ref{eq-mov-CR2D})) and with a one-dimensional one in the second case (Eq~(\ref{eqMoG-CR2})). We see that the overall  behavior of the CRLB along the drifting direction for a moving source is fairly similar in the two cases. It follows that most of the results and conclusions of Sect.~\ref{sec:StudyCR_1D} can be extended for observations performed with two-dimensional detectors.

In particular, the same distinct areas remain visible: the oversampled area where the CRLB decreases when pixel size increases, the well-sampled area where the CRLB remains constant (for pixel sizes close to the FWHM half-value), the intermediate area where the CRLB oscillates (for values of the pixel size between the FWHM and the drifting parameter $L$), and the undersampled area where the CRLB increases when the pixel size increases (for a
pixel size $> 3.0$" not shown in Fig.~(\ref{figmov_2D})).

Concerning the impact of pixel decentering on the CRLB behavior, the two-dimensional case is similar to the linear case: in the oversampled and well-sampled areas, the pixel decentering effect is almost negligible, while in the intermediate and undersampled areas, the CRLB behavior is strongly affected by this effect. However, the specific pixel size values of the intermediate regime where the CRLB values were almost not affected by the decentering effect seen in the one-dimensional case (as for a pixel size value of $1.3$" for the top source in Fig.~\ref{figbcas3}) are more affected by the decentering effect in the two-dimensional case. The reason is that with a two-dimensional detector, the pixel decentering of the PSF has not one, but two degrees of freedom, which are represented by two offsets along the X- and Y-axes, and there is no reason for these two decentering components to be counterbalanced by the drift of the source for the same pixel size. There is one exception, however: when the inclination of the DPSF with respect to the X axis of the CCD frame equals $45^o$, the decentering along both axis vanishes for the same pixel size values (see, for instance, the bottom curve of Fig.~\ref{figmov_2D} at the pixel size of $1.45$"). This configuration can be particularly advantageous when an observation of a moving target has to be planned in an undersampled regime.  

Finally, we also note, as was the case for the linear detector case, that the minimum value reached by the CRLB is not in the undersampled area of the plot, and that the combined effect that
is due to decentering and orientation of DPSF can be neglected in the oversampled and well-sampled areas.\\
   
\underline{Oversampled case approximations:}

We can produce accurate approximations of Eq.~(\ref{eq-mov-CR2D}) for
a two-dimensional detector that observes in an oversampled regime by using
$\overline{\Phi}_{M^2}$ as the normalized PSF in Eq.~(\ref{eqLB-2D})
(when the background dominates the total flux), and in
Eq.~(\ref{eqLF-2D}) (when the source dominates the total flux) and by
replacing the derivatives with respect to $x_c$ by the derivatives
with respect to $u_c$. Moreover, following a remark of
Sect.\ref{subsec:CR_GEN_2D}, we can compute in the $U\times V$-space
instead of the $X\times Y$-space the two integrals of
Eqs.~(\ref{eqLB-2D}) and (\ref{eqLF-2D}) (whose reciprocals are
denoted $I_{M_1^2}^U$ and $I_{M_2^2}^U$ , respectively).

The approximations of these reciprocals are given in
Appendix~\ref{app:MsL} (when the drifting parameter is small) and in
Appendix ~\ref{app:MbL} (when the drifting parameter is large). When the drifting parameter is small ($L\leq1.5\times$FWHM),
expressions (\ref{eqLB-2D}) and (\ref{eqLF-2D}) for the CRLB along the
drifting direction become
\begin{eqnarray}
\label{eqLB2_2Dm}
\sigma_{CR_U}^{2} &=& \Frac{8\pi \tilde B\sigma^4}{\tilde F ^2
  ~ {\Delta x}^2} ~\left[1 +
  \frac{1}{2}\left(\overline{L}\right)^2 +
  \frac{1}{12}\left(\overline{L}\right)^4
+o\left[\overline{L}^{8}\right]\right]\\
&&~~~~~~~~~~~~~~~~~~~~~~~~~~~~~~~~~~~~~~~~~~~~~~~~~~~~\mathrm{if}~~\tilde{F}/\tilde{B}<<1
\nonumber\\
\label{eqLF2_2Dm}
\sigma_{CR_U}^{2} &=& \Frac{\sigma^2}{\tilde F} ~ \left[ 1 +
  \frac{1}{3}\left(\overline{L}\right)^2
+o\left[\overline{L}^{8}\right]\right]~~~~~~~~~~\mathrm{if}~~\tilde{F}/\tilde{B}>>1
,\end{eqnarray}
respectively, while when the drifting parameter is large ($L\geq 2\times$FWHM), the expressions become
\begin{eqnarray}
\label{eqLB3_2Dm}
\sigma_{CR_U}^{2} &=& \Frac{8\pi \tilde B\sigma^4}{\tilde
  F^2~{\Delta x}^2}~\overline{L}^2
~~~~~~~~~~~~~\mathrm{if}~~\tilde{F}/\tilde{B}<<1\\
\label{eqLF3_2Dm}
\sigma_{CR_U}^{2} &=& 1.11 ~\Frac{\sigma^2}{\tilde F} ~ \overline{L} 
~~~~~~~~~~~~\mathrm{if}~~\tilde{F}/\tilde{B}>>1
.\end{eqnarray}
Here we recall that the normalized drifting parameter
$\overline{L}$ is equal to $L/2\sigma$ and that the background $\tilde
B$ follows the two-dimensional expression (\ref{eqBG3}) and not
Eq.~(\ref{eqBG1}) that was used in the linear case.

We first observe that when the source dominates the total flux,
expressions (\ref{eqLF2_2Dm}) and (\ref{eqLF3_2Dm}) are identical to
expressions (\ref{eqLF2}) and (\ref{eqLF3}), respectively, that
were used in Sect.~\ref{subsec:StudyCR_CSTBGperDEG_OScase} for
the source that was observed in the same regime with a linear detector. When
the background dominates the total flux, expressions (\ref{eqLB2_2Dm})
and (\ref{eqLB3_2Dm}) can be deduced from approximations
(\ref{eqLB2}) and (\ref{eqLB3}) for a linear detector by multiplying
them by $2\sqrt{\pi}\sigma/\Delta x$.

Thus, all the remarks concerning the behavior and the precision of
the CRLB approximations given for a linear detector in
Sect.~\ref{subsec:StudyCR_CSTBGperDEG_OScase} can be trivially
extended for a two-dimensional detector. In particular, and in a
consistent way, for a slowly moving source, the limit of
the CRLB expressions (\ref{eqLB2_2Dm}) and (\ref{eqLF2_2Dm}) when $L$
approaches zero are equal to expressions
(\ref{eqLB2_2DG}) and (\ref{eqLF2_2DG}), respectively, of a stationary source in the
same regime.  Similarly, when $L$ increases, the four CRLB
approximations of this section are logically degraded compared to the
CRLB expressions for a stationary source in the same regime.\\

\underline{Optimum exposure time:} 

We recall that in Sect.~\ref{subsec:OptExpTime} we defined the
{\it{optimum exposure time}} as the exposure time that minimizes the
CRLB for a moving source, and we distinguished the optimum exposure
time for an isolated image from the optimum exposure time of one image
contained in a set of N images with a fixed total duration $T_a$.

Concerning the optimum exposure time for an isolated image, all the
results obtained in that previous subsection were based on
approximations of the CRLB in the oversampled regime. In this
regime, the CRLB expressions for a moving source observed with a
two-dimensional detector and with a linear one differ only by a
  scale factor that is independent of time, therefore all the results of
Sect.~\ref{subsec:OptExpTime} can be easily generalized to the
two-dimensional case. Nevertheless, for a two-dimensional
detector array, the background flux $\tilde{B}$, instead of obeying
Eq.~(\ref{eqBG2}), follows Eq.~(\ref{eqBG4}),\begin{equation}
\tilde{B}=b_1 ~ T_{e} + b_0 = b_s ~ T_{e} ~ {\Delta x}^2 + d ~ T_{e}+RON^2
\label{eqBG4}
.\end{equation}
The unit of the sky component ($b_s$) is now $e^-$/arcsec$^2$/sec.

In particular, from a quantitative point of view, the expression for
the ultimate lower limit for the astrometric precision of a moving
source (\ref{eqLCRTe}), as well as the expression of the lower limit
$T_o$ for the optimum exposure time (\ref{eqTo}), remain unchanged in the
two-dimensional case. Similarly, expression (\ref{eqLFo}) of the CRLB
when the source dominates the total flux for an exposure time equal to
$T_o$ is still valid in the two-dimensional case, while expression
(\ref{eqLBo}) of the CRLB when the background dominates the total flux has
to be multiplied by a factor $2\sqrt{\pi}\sigma/\Delta x$ for the
two-dimensional detector array.

Similar as in the one-dimensional case, if the lower limit $T_o$ is used as exposure time and if the RON is negligible, the degradation of the CRLB for a moving source along the drifting direction is therefore larger by $19\%$ and $33\%$  than the CRLB for the same source when it is observed with no motion and during the same exposure time. When the RON is not a negligible part of the total background, the CRLB degradation caused by the motion of the source will be slightly larger. When the total flux is dominated by the source, the CRLB of a
moving source can reach values slightly below the value given by expression (\ref{eqLFo}), but the maximum possible improvement is $7\%$ and only for an (ideal) measurement without background and with an exposure time equal to infinity.

Concerning the optimum exposure time $T_N$ for a whole set of $N$ images with a fixed total duration $T_a$, the method applied for a linear detector is directly applicable for a two-dimensional detector by using Eq.~(\ref{eq-mov-CR2D}) to model $\sigma_i$ (instead of Eq.~(\ref{eqMoG-CR2})) in expression (\ref{eqRatioCR}) of the CRLB ratio. A binary search algorithm applied to the localization of the maximum of this new CRLB ratio expression allows us to compute $T_N$ by optimizing the CRLB of the set of images observed now with a two-dimensional detector array.

\subsubsection{CRLB along the direction normal to the motion}
\label{subsubsec:CR_mov_2D_V}

For moving sources observed with a two-dimensional detector, 
the behavior of the CRLB along the direction normal to the motion is not adequately
represented by the expressions developed for stationary sources (and
this even though the cross-section of the DPSF along this direction is
similar to the PSF of stationary sources).
In the two-dimensional case, the measurement of the CRLB along
the direction normal to the motion indeed depends on the number of pixels
covered by the PSF along the direction perpendicular to this
measurement (i.e., the direction of the motion). An increase in the
underlying number of pixels in the direction of motion decreases
the S/N without adding any information for the PSF centering along the
direction normal to the motion.  Since the underlying number of pixels
along the direction of motion is directly related to the source-drifting parameter $L$, the expression of the CRLB along the direction
normal to the motion should depend on $L$ as well. All other things
remaining equal, the smaller $L$, the higher the S/N and the smaller
the CRLB along the direction normal to the motion. The
lowest value of the CRLB in this case would be reached when the
drifting parameter $L$ equals zero (i.e., for the stationary source).
       
The generic expression for the CRLB along the direction normal to the
motion (i.e., along the $V$-axis) can be deduced from expression
(\ref{eqcr2D-1}) by replacing $\overline{\Phi}$ by
$\overline{\Phi}_{M^2}$ and the derivative with respect to $x_c$ by
the derivative with respect to $v_c$.  Then, the expression of the
CRLB along the direction normal to the motion is the following:
\begin{equation}
\sigma_{CR_V}^{2} =\Frac{8\pi\sigma^6 L^2}{\tilde{F}^2}~\Frac{1}{\sum_{j=1}^{n_y}\sum_{i=1}^{n_x}\frac{\left(N_i^j\right)^2}{D_i^j}}
\label{eq-mov-CR2D_v}
,\end{equation}
where now
\begin{eqnarray}
\begin{array}{rcl}
N_i^j &=& \int_{y_j^-}^{y_j^+} \int_{x_i^-}^{x_i^+}
\left(v-v_c\right)e ^{-\frac{1}{2}\left(
  \frac{v-v_c}{\sigma}\right)^2}\left(P(U_2)-P(U_1)\right) dx dy\\
D_i^j &=& \tilde B + \frac{\tilde{F}}{2\sqrt{2\pi}\sigma L} I_i^j\\
I_i ^j&=&\int_{y_j^-}^{y_j^+} \int_{x_i^-}^{x_i^+} e
^{-\frac{1}{2}\left(
  \frac{v-v_c}{\sigma}\right)^2}\left(P(U_2)-P(U_1)\right) dx dy.
\end{array}\nonumber 
\end{eqnarray}
For similar reasons as those given in the case of the CRLB along the drifting direction, in expression (\ref{eq-mov-CR2D}), the explicit integrals cannot
be easily avoided here.

As expected, we see from Eq.~(\ref{eq-mov-CR2D_v}) that the CRLB expression along the direction normal 
to the motion depends of the source-drifting parameter (through $U_1$ and $U_2$ as defined in Sect.~\ref{sec:SpreadFun}).\\

\underline{Oversampled case approximations:} 

We can give accurate approximations for the CRLB along the direction
normal to the motion by using the 2D MoG function
$\overline{\Phi}_{M^2}$ and by now substituting the partial derivative
with respect to $x_c$ by the one with respect to $v_c$ in
Eqs.~(\ref{eqLB-2D}) and (\ref{eqLF-2D}). The approximations of the
reciprocals of the two integrals involved in Eqs.~(\ref{eqLB-2D}) and
(\ref{eqLF-2D}) are given in Appendix~\ref{app:MsL} (for a small
drifting parameter) and in Appendix ~\ref{app:MbL} (for a large
drifting parameter). The reciprocals of the integrals involved in
Eqs.~(\ref{eqLB-2D}) and (\ref{eqLF-2D}) are denoted $I_{M_1^2}^V$ and
$I_{M_2^2}^V$ , respectively.

First, when the source dominates the total flux, the function $I_{M_2^2}^V$ is independent of the drifting parameter $L$ and equals $I_{S_2^2}$ , which corresponds to the same integral computed with a normalized PSF corresponding to a stationary source observed with a two-dimensional detector (see Appendix~\ref{app:MsL} and Sect.~\ref{subsec:CR_stat_2D})). In the oversampled case, when the background component is insignificant in comparison to the source flux, we see that the CRLB of a moving source measured along the direction normal to the motion is identical to the CRLB of the corresponding stationary one and the CRLB expression is given by Eq.~(\ref{eqLF2_2Ds}),
\begin{eqnarray}
\label{eqLF2_2Ds}
\sigma_{CR_{V}}^{2} &=& \Frac{\sigma^2}{\tilde F} ~~~~~~~~~~~~~~~~~~~~~~~~~~~~~~~~~~~~\mathrm{if}~~\tilde{F}/\tilde{B}>>1
.\end{eqnarray}

Second, when the background dominates the total flux, the function
$I_{M_1^2}^V$ depends on the source-drifting parameter $L$ through a
scale factor that involves the integral over the whole space of the square
of the 1D MoG function $\overline{\Phi}_M$
(Appendix~\ref{app:MsL}). As previously, to compute
this integral analytically, we have to distinguish two cases: slow-moving sources,
and fast-moving sources (see Appendix~\ref{app:MsL} and
\ref{app:MbL} for details), and the corresponding CRLB expressions
become
\begin{eqnarray}
\label{eqLB2_2Ds}
\sigma_{CR_V}^{2} &=& \Frac{8 \pi \tilde B\sigma^4}{\tilde F ^2 ~
  {\Delta x}^2}  ~ \left[1 + \frac{1}{6}\left(\overline{L}\right)^2
  - \frac{1}{180}\left(\overline{L}\right)^4
  +o\left[\overline{L}^{6}\right]\right]\\
&&~~~~~~~~~~~~~~~~~~~~~~~~~~~~~~~~~~~~~~~~~\mathrm{if}~
L\leq1.5\times\text{FWHM}\nonumber\\
\label{eqLB3_2Ds}
\sigma_{CR_V}^{2} &=& \Frac{8\pi\tilde B\sigma^4}{\tilde F ^2 ~
  {\Delta x}^2}
~ \Frac{\overline{L}^2}{(\overline{L}\sqrt{\pi}-1)}
~~~~~~~~~~\mathrm{if}~L \geq2\times\text{FWHM}
,\end{eqnarray}

respectively. As expected, the limit of the CRLB expression (\ref{eqLB2_2Ds}) when
$L$ approaches zero is equal to the CRLB expression (\ref{eqLB2_2DG})
of a stationary source in the same regime. When $L$ increases,
expressions (\ref{eqLB2_2Ds}) and (\ref{eqLB3_2Ds}) are degraded
compared to the CRLB expression of a stationary source, but less so
than the CRLB measured along the drifting direction as represented by
expressions (\ref{eqLB2_2Dm}) and (\ref{eqLB3_2Dm}), respectively.

When we substitute in Eq.~(\ref{eqLB2_2Ds}) $\overline{L}$ by the normalized optimal drifting parameter $\overline{L}_o$ , we obtain the minimum degradation of the CRLB along the direction normal to the motion for the case when the background dominates the total flux.  We obtain
\begin{eqnarray}
\label{eqLBo_2Dm_V}
\sigma_{o_V}^{2} = \Frac{8\pi \tilde B\sigma^4}{\tilde F ^2
  ~ {\Delta x}^2} ~ \left[ 1.20 + 0.27\mu_b - 0.38\mu_b^2 \right] ~~\mathrm{if}~\tilde{F}/\tilde{B}<<1&&
,\end{eqnarray}
where $\mu_b$ is now computed with $b_0$ and $b_1$ of Eq.~(\ref{eqBG4}).

By using Eq.~(\ref{eqLBo_2Dm_V}), we conclude that if the optimum exposure time $T_o$ is used as exposure time, the maximum degradation of the CRLB along the direction normal to the motion is $10\%$ larger than the CRLB of the corresponding stationary source when the RON is negligible and slightly larger if it is not. For a bright sources, this degradation tends to zero, as shown by Eq.~(\ref{eqLF2_2Ds}).

\begin{figure}[ht]
\begin{center}
\includegraphics[width=9cm]{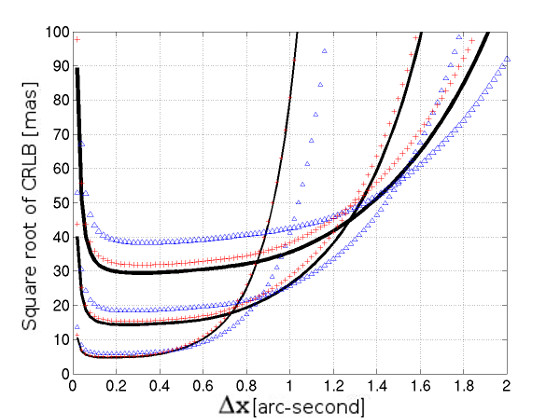}
\caption{\textit{Square root of the CRLB in mas as a function of pixel size $\Delta x$
    for sources observed with a two-dimensional array detector when
    the background flux per pixel is given by Eq.~(\ref{eqBG3})
    ($b=6000$ $e^-/$arcsec$^2$, $D=0~e^-$, $RON=5~e^-$ , and $F=6000$
    $e^-$). The solid lines (thin, normal, and bold) are computed with
    expression (\ref{eq-stat-CR2D}) for a stationary source with an
    image quality FWHM of $0.5$, $1.0$, and $1.5$" , respectively (and all
    centered on a given pixel). The lines with a plus and with
    triangles correspond to the same curves, but in the moving case
    when the drifting parameter is optimum $(L=L_o)$. The lines with a plus and with triangles
    correspond to the CRLB along the direction normal and parallel to the drifting motion, respectively.}}
\label{fig2DF}
\end{center}
\end{figure}

To conclude this section and to summarize the main results obtained
for the limit of the astrometric precision for moving sources observed
with a two-dimensional detectors, we present in Fig.~\ref{fig2DF} the
CRLB behavior for a circularly symmetric source in the stationary
case and in the moving case.  For moving sources, we consider that the
exposure time is the optimum exposure time presented previously, which
means that the drifting parameter equals the optimum drifting
parameter $L_o$.

All other things being equal, we represent in Fig.~\ref{fig2DF} (for
three different values of image quality FWHM) the behavior of the
CRLB in the stationary case (based on Eq.~(\ref{eq-stat-CR2D})) and
the behavior of the CRLB along two directions in the moving case:
along the drifting direction (based on Eq.~(\ref{eq-mov-CR2D})), and
along the direction normal to the motion (based on
Eq.~(\ref{eq-mov-CR2D_v})).

We see that with this optimum exposure time, the CRLB overall
behavior is very similar in both the moving and the stationary
case.  In particular, the CRLB oscillations visible in the
intermediary-sampled regime for moving sources are not present here
(this is due to the choice of exposure time, which leads to a drifting
parameter below the FWHM, while this intermediary regime only appears for
drifting parameters larger than the FWHM).

For the same FWHM, the CRLB minimum values for stationary as well as
for moving sources in Fig.~\ref{fig2DF} are reached in the
well-sampled area. The CRLB minimum value for a stationary source is
always lower than that of a moving source. However, the minimum value
for the CRLB along the direction normal to the motion is noticeably
closer to the stationary one than the CRLB along the drifting
direction: for the three sources in Fig.~\ref{fig2DF}, this
degradation of the CRLB minimum value amounts to around $5\%$ along
the direction normal to the motion and to around $25\%$ along the
drifting direction. Even though the minima of the CRLB are reached for
relatively large pixel sizes (close to one-third of the FWHM), these
percentages are in full agreement with the interval of degradation
deduced from the CRLB approximations in the oversampled case when the
exposure time equals $T_o$ and when the RON is negligible
(i.e., between $19\%$ and $33\%$ for the CRLB along the drifting
direction, and between $0\%$ and $10\%$ for the CRLB along the
direction normal to the motion).  This also visually confirms that the
estimators developed in the oversampled case for slow-moving sources
still give a good approximation in the well-sampled case (the
estimator for CRLB along the drifting direction is given by the
maximum of the two expressions (\ref{eqLB2_2Dm}) and
(\ref{eqLF2_2Dm}), while the estimator for the CRLB along the
direction normal to the motion is given by the maximum of the two
expressions (\ref{eqLB2_2Ds}) and (\ref{eqLF2_2Ds})).

\section{Comparison with astronomical observations.}
\label{sec:CompWithImages}
In the previous sections, several theoretical predictions have been derived for the limit of the astrometric precision that
can be reached for both stationary and moving sources that are observed with digital-detector arrays. For stationary sources, the cornerstone of our analysis is Eq.~(\ref{eq-stat-CR2D}), which gives the expressions for the CRLB along any direction. For moving sources, our results are based upon the two Eqs.~(\ref{eq-mov-CR2D}) and (\ref{eq-mov-CR2D_v}), which give the expression of the CRLB along the direction parallel and normal to the source drift, respectively. In this section, the theoretical results derived using these equations are compared to the results of astrometric reductions performed on simulated and real astronomical observations of stationary and moving sources detected with CCD-sensors.

The astrometric reduction process encompasses the detection of all 
sources in an image above a specified threshold, determining for each
source their $X$ and $Y$ photocenter
positions, their total flux, and the value of the image quality FWHM
by fitting a 2D MoG function with an unweighted
  least-squares algorithm (LSA hereafter). The drifting parameters $L$
and the angle $\alpha$ of the MoG functions are assumed to be
known. These tasks are performed by using the GBOT astrometric
reduction pipeline \citep{Bouquillon14}, which is one of the tools
developed for the GBOT satellite tracking project of ESA's Gaia spacecraft, see
Sect.~\ref{subsec:CompWithRealData}. For each determination of
a stellar photocenter, the GBOT reduction pipeline also provides two
estimates, $\sigma_U$ and $\sigma_V$, that correspond to the
astrometric precision along the drifting direction and along the
direction normal to the motion, respectively. These estimates are the
standard deviations provided by the LSA through the PSF fitting
process. In the next subsection we use simulated astronomical images
to first demonstrate the quality of these estimators, and
second, to compare the CRLB expressions of previous sections with
the astrometric precision achieved by the GBOT image analysis.

\subsection{Comparison with simulated data.}
\label{subsec:CompWithSimData}

\begin{figure}[ht]
\begin{center}
\includegraphics[width=9cm]{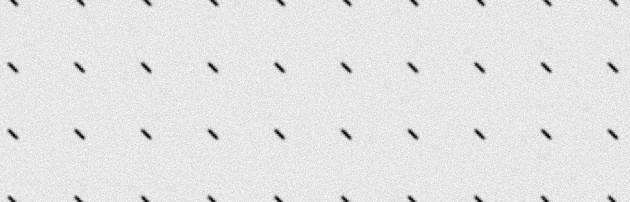}
\caption{\textit{Portion of one of the $63$ simulated images (for $L=4$~FWHM).}}
\label{figImaSim}
\end{center}
\end{figure}

Three sets of $21$ simulated images with $100$ drifting sources in each image were created.  An example of these images is given in Fig.~\ref{figImaSim}.
All the objects of a given image are similar (motion, FWHM, total flux, background, etc.). In a given set, the only difference between the objects of two distinct images are their drifting parameter $L$ (which varies, taking 21 different values between zero and ten times the FWHM). The total flux of each object in set $S_1$ is $30000~e^-$, in set $S_2$ it is $8000~e^-$ , and in set $S_3$ it is $4000~e^-$. For all the images, the background mean value is $100~e^-/pix$, the pixel size is $0.3$", and the FWHM is $4.7pix$. At the end of this image creation process, we added noise following a Poisson distribution. The peak S/N of the objects of sets $S_1$, $S_2,$ and $S_3$ are $33$, $15,$ and $10,$ respectively, in the stationary case and $8.4$, $2.9,$ and $1.5,$ respectively, when $L=10$~FWHM.

\begin{figure}[ht]
\begin{center}
\includegraphics[width=9cm]{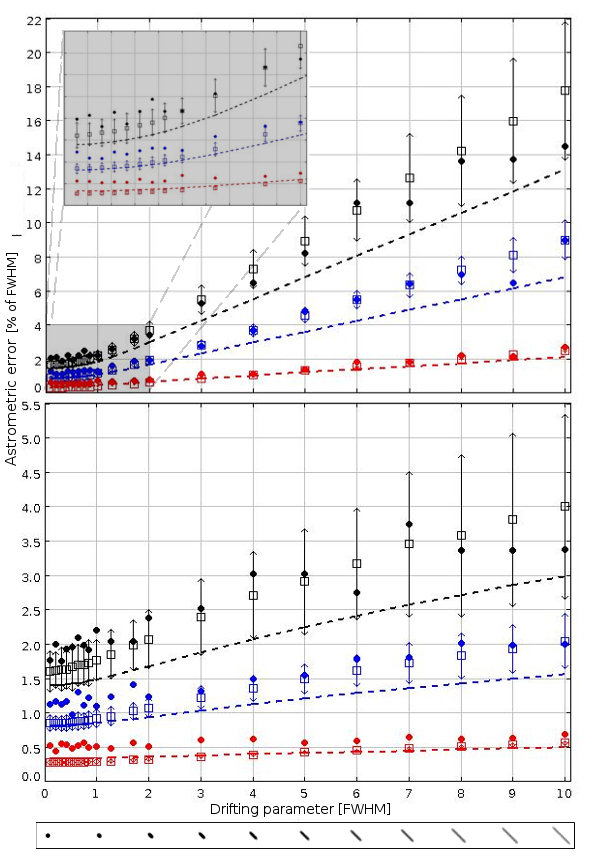}
\hspace{0.5cm}
\caption{T\textit{rue astrometric errors}, their GBOT estimates,
  and the corresponding CRLB values for different values of the
source-drifting parameter between zero and $\text{ten
times the }$FWHM (along the
  $U$-axis in the top panel, and along the $V$-axis in the bottom
  panel). Each filled circle is the standard deviation of the
  "true astrometric errors" of one simulated image. Each open
  square and related error bar are the mean value and standard
  deviation of the GBOT estimates of the astrometric uncertainties.
  The dashed curves are the corresponding value of the CRLB computed
  with Eqs.~(\ref{eq-mov-CR2D}) and (\ref{eq-mov-CR2D_v}) for the top
  and bottom panels, respectively. The lower curves are for bright
  sources of set $S_1$, the medium curves for sources of set $S_2$,
  and the upper curves for faint sources of set $S_3$ (see the details
  concerning the characteristics of each set in the text).}
\label{fig_mogQ}
\end{center}
\end{figure}

The astrometric reduction of all these simulated images is performed with the GBOT pipeline, as explained in the introduction of this section. The results are summarized in Fig.~\ref{fig_mogQ} for fast- and slow-moving sources. 
This figure shows the astrometric errors, their GBOT estimates, and the corresponding CRLB values (in percent of the FWHM) with respect to the drifting parameter $L$
(in units of the FWHM): the top panel shows the astrometric error along the direction of motion ($U$), while the bottom panel shows the astrometric error along the direction perpendicular to the motion ($V$). The lower curves are for the sources of set $S_1$, the medium curves for the sources of set $S_2$, and the upper curves are for the sources of set $S_3$.

Each filled circle depicts the standard deviation of the differences between the known photo-centers and those estimated by the GBOT reduction pipeline for the $100$ similar objects of one simulated image. Each open square and each corresponding error bar show the mean value and three times the standard deviation, respectively, of the astrometric uncertainties estimated by GBOT for the $100$ objects of one image. These figures show that the estimates provided by the GBOT pipeline are quite robust: the majority of the true astrometric errors (filled circles) are within the error bars derived from the GBOT estimates. It seems
that the quality of the estimates is slightly poorer for the brightest sources (bottom curves) or when the true astrometric error is very small (below $1\%$ of the FWHM).

For a drifting parameter $L$ close to four times the FWHM (similar to the case of the moving objects in the real image used in the next subsection), we see from Fig.~\ref{fig_mogQ} that the GBOT estimates are quite robust, and we observe that only for the brightest sources the estimate of the astrometric error along the direction normal to the motion overestimates the true astrometric
error by about $30\%$ . This is most likely related to the loss of optimality (in the CRLB sense) of the LSA parameter estimating method, as recently 
demonstrated by \citet{Lobos15} (see especially their Eq.~(26) in proposition 3, and their Fig.~4).

For each simulated image we can also compute the corresponding CRLB values along the drifting direction $U$ (with the help of Eq.~(\ref{eq-mov-CR2D})) and in the direction $V$ normal to the motion (with the help Eq.~(\ref{eq-mov-CR2D_v})). The dashed lines correspond to these CRLB values, which are, from a theoretical point of view, the lower limits of the true astrometric errors, regardless of the centroiding method used. We note here that it can indeed be 
proven that the CRLB is theoretically unreachable in the context of digital detectors with Poisson noise, as demonstrated by \citet{Lobos15} (see Sect.~3.1, ``Nonachievability'' and, in particular, their proposition 2).

As expected, we see that the theoretical CRLB values (dashed lines) are below the true astrometric errors (filled circles) computed from the centroiding performance of the GBOT reduction pipeline in all the cases. We also observe that the degradation of the GBOT true astrometric errors with respect to the drifting parameter $L$ is fully consistent with the CRLB overall expected behavior. Finally, we note that the differences between the two are small, but that improvements of the centroiding method accuracy are still possible (especially when the true astrometric error is small, below $1\%$ of the FWHM). We are currently working on implementing an algorithm within GBOT that is based on a maximum likelihood estimate, which we expect it will render better results when the S/N is high (\citet{Lobos15}, see especially their Fig.~8), this will be reported in a future paper.

In summary, our numerical results demonstrate the adequacy of the centroiding performance of the GBOT reduction pipeline (since the observed accuracies are close and in good agreement with the CRLB). Reciprocally, we validate the CRLB expressions (\ref{eq-mov-CR2D}) and (\ref{eq-mov-CR2D_v}) in the cases of moving sources observed with a two-dimensional array. They also demonstrate the robustness of the astrometric centroiding uncertainty estimate provided by GBOT. This was a necessary step before we proceed to the
comparison of the CRLB expressions with true astronomical images of moving sources because in the real case, we do not have access to the true astrometric errors, but only to the GBOT estimates of astrometric uncertainties.

\subsection{Comparison with real images.}
\label{subsec:CompWithRealData}

\begin{figure}[ht]
\begin{center}
\includegraphics[width=9cm]{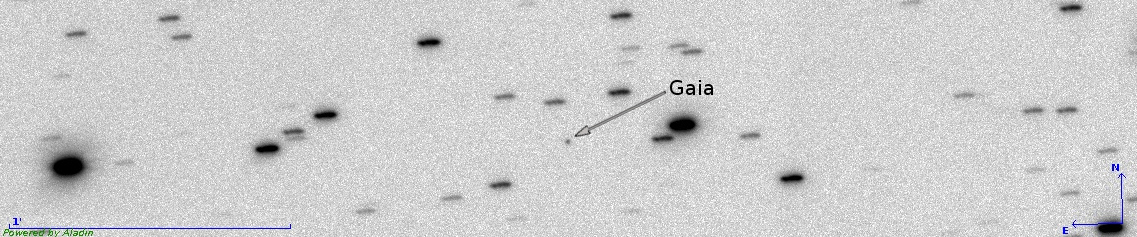}
\caption{\textit{Portion of one GBOT image taken with the 12th CCD of
    the OmegaCam camera of VST on January 6, 2014. The point-source
    at the center of the image is the satellite Gaia, and all the
    others sources are the background stars, elongated because of
    their speed relative to Gaia.}}
\label{figImaVST}
\end{center}
\end{figure}

For this aim, we selected astronomical observations performed with the
VLT Survey Telescope (VST) of the European Southern Observatory (ESO), which is a 2.6 m wide-field optical survey
telescope located on the VLT platform at Cerro Paranal, Chile. This
telescope is equipped with one focal plane instrument, OmegaCam, a
large-format (16k$\times$16k pixels) CCD mosaic camera with a large
corrected field of view of $1^o\times 1^o$. The observations were
obtained in the framework of the GBOT campaigns (see \citet{Altmann14}
or \citet{Jordan13} for more details). The aim of these campaigns is to track the Gaia
satellite to improve the accuracy of its orbit determination. Since
the Gaia satellite is faint and moves relative to the reference
frame defined by the background stars, the telescope tracking is
locked on the Gaia speed. This tracking mode allows recording Gaia as
a stationary source in the CCD frame (allowing a better precision of
the photocenter determination, as extensively shown in this paper). On
the other hand, this means that the - stationary - stars are moving
sources in this frame, and their images on the CCD are elongated (see
Fig.~\ref{figImaVST}).

\begin{figure}[ht]
\begin{center}
\includegraphics[width=9cm]{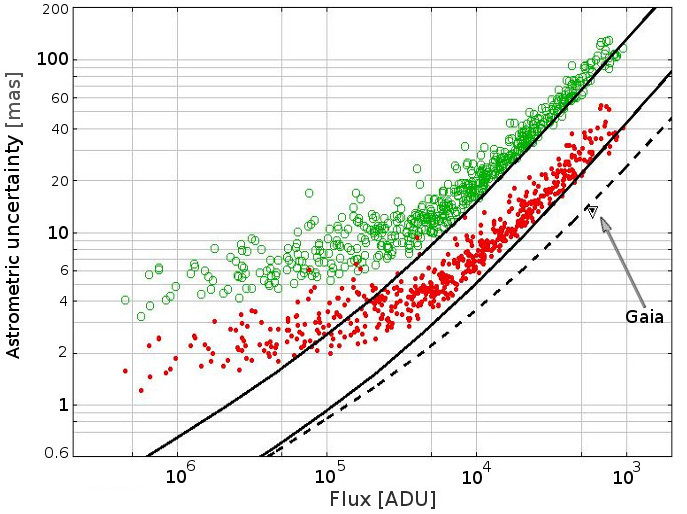}
\hspace{0.5cm}
\caption{Astrometric uncertainty estimates for the photocenters of stars
  (and Gaia) as measured by the GBOT reduction pipeline on the VST
  image as a function of flux (in ADU). The open and  filled circles are the uncertainty estimates along the $U$ and the $V$
  directions, respectively (the open and filled diamonds are for
  Gaia). The dashed curves correspond to the corresponding CRLB values
  for stationary sources, while the solid curves are the CRLB values
  for moving sources. The CRLB has been estimated with the expressions developed for
  two-dimensional array detectors (Eq.~(\ref{eq-stat-CR2D}) for
  stationary sources and Eqs.~(\ref{eq-mov-CR2D}) and
  (\ref{eq-mov-CR2D_v}) for moving sources for the CRLB along the $U$
  and $V$ directions, respectively).}
\label{fig_CRapp1}
\end{center}
\end{figure}

The image selected here has been taken at the very beginning of the
mission, before Gaia settled in its operational orbit around Sun-Earth
L2.  The reason for this choice is that the apparent mean speed of
Gaia $V_G$ after arrival at its final location near L2 as seen from Earth at night is usually around 0.02"/sec, while in this image
the speed of Gaia is about 0.063"/sec. Thus, these data allow us to
study the behavior of faster sources than the observations performed
after Gaia settled in its final orbit.

In this image, the mean quality FWHM is $1.00$".
The exposure time $T_e$ was $60$ seconds, Gaia is stationary with
respect to the CCD-frame, and the star's motion with respect to the
CCD frame is of the same amplitude and opposite to the motion of Gaia in
relation to the star frame. This means that the stellar~motions are oriented with
an angle $\alpha=-9.73$ degrees with respect to the $X$-axis of the
CCD, and their drifting parameter $L$ is $3.81$" (which corresponds
to a normalized drifting parameter $\overline{L}$ of $4.45$).  The
mean standard deviation of the image background noise is
$8.71$~ADU/pix. The detector has negligible dark-current and
$RON$. Its gain is $2.81~e^-/ADU$. The pixel size is $0.214$" (lower
than one-third of the FWHM), which places this image between the
oversampled and the well-sampled regime.

The great advantage of this image (with the stars being the ``moving
sources'') for this particular study is that it provides many sources
with very different S/N, but with the same apparent motion (same
apparent speed and direction) and the same instrumental and
meteorological conditions (same image quality FWHM, same background
noise, etc.).

In the same way as for the simulated images presented in the
previous subsection, we performed the star centroiding determination
by using the GBOT astrometric reduction pipeline with the MoG
function as DPSF. As mentioned previously, the drifting parameters $L$
and angle $\alpha$ of the MoG function are assumed to be
known. Their values are those mentioned above and are deduced from
the ephemeris of Gaia. To avoid including astronomically extended
objects in our analysis, we only keep those objects detected by the
GBOT reduction pipeline for which the fitted FWHM
equaled the mean image seeing with a margin of error of $25\%$. For
Gaia, which is a stationary source in this image, we fit a circularly
symmetric two-dimensional Gaussian.

The astrometric results of the reduction described in the previous
paragraph are presented in Fig.~\ref{fig_CRapp1}. This figure shows
the GBOT estimates of the astrometric uncertainties $\sigma_U$ and $\sigma_V$ for
all the stars (and for Gaia) as a function of flux. The open and
filled circles are for stellar $\sigma_U$ and $\sigma_V$ , respectively
(the open and filled diamonds are for Gaia's $\sigma_U$ and
$\sigma_V$). As expected, we see
that all else being equal, the brighter the source, the smaller the two uncertainty estimates for $\sigma_U$ and $\sigma_V$ of the
star positions along the $U$ and $V$ axes. We then note that for every 
moving star, $\sigma_U$ is always larger than $\sigma_V$, which means
that the degradation of the astrometric precision is larger along the
drifting direction than along the direction normal to the motion, as
we discussed in the previous sections. For Gaia - the stationary
source - the two uncertainty estimates $\sigma_U$ and $\sigma_V$ have the
same value since a circular symmetric Gaussian has been fit.
For stars with fluxes similar to that of Gaia, the uncertainty estimates of
drifting sources (i.e., stars) are clearly larger than those of Gaia
(whether it is along the drifting direction or perpendicular to
it). Finally, we observe two distinct regimes for $\sigma_U$ in
comparison to $\sigma_V$: for bright sources the astrometric error is
slowly degraded when the flux decreases, while for faint sources this
degradation is faster. For this image, the bifurcation between these
two regimes is for a source flux of around $20\,000$~ADU (which
corresponds to an S/N~$\sim100$).

We now compare the uncertainty estimates of this image with the CRLB expressions  presented in the previous section. The bold and thin solid lines plotted in the same figure are computed by using expressions (\ref{eq-mov-CR2D}) and (\ref{eq-mov-CR2D_v}) of the CRLB along the $U$ and $V$ directions, respectively. The dashed line is based on Eq.~(\ref{eq-stat-CR2D}) of this paper, representing the CRLB of a stationary source observed with a two-dimensional array detector. We see, as expected, that the bold solid lines represent the lower limit of the $\sigma_U$ of the drifting sources (open circles), while the thin solid line is the lower limit of the $\sigma_V$ of the drifting source (filled circles). We also see that in the regime of faint sources, the difference between CRLB expressions and the uncertainty estimates is very small. Similarly, for equivalent flux values, the difference between the uncertainty estimate of the stationary source (Gaia) and the uncertainty estimates $\sigma_V$ along the direction normal to the motion of the drifting sources (stars) is clearly explained by the
difference between expressions (\ref{eq-stat-CR2D}) and (\ref{eq-mov-CR2D_v}).

In the regime of bright sources, we observe from Fig.~\ref{fig_CRapp1}
that the differences between the CRLB expressions and the GBOT uncertainty
estimates increase with the rise of the source brightness (visible
in the left part of Fig.~\ref{fig_CRapp1}). This is
likely due to two effects that should be studied in more detail, namely, (a)
the choice of the LSA method, which is well-adapted for estimating
source parameters for a low S/N (as is the case of the
target, Gaia), but not for bright sources, as explained in detail in
\citet{Lobos15} (see also Sect.~\ref{subsec:CompWithSimData}), and (b)
because of a difference between our adopted DPSF fitting processing and
the CRLB theory developed in this paper. We not only estimate
one coordinate of the source photocenter (as assumed in the
theoretical sections of this paper), but four parameters: the two
coordinates of the photocenter, the total flux, and the FWHM.

\begin{figure}[ht]
\begin{center}
\includegraphics[width=8cm]{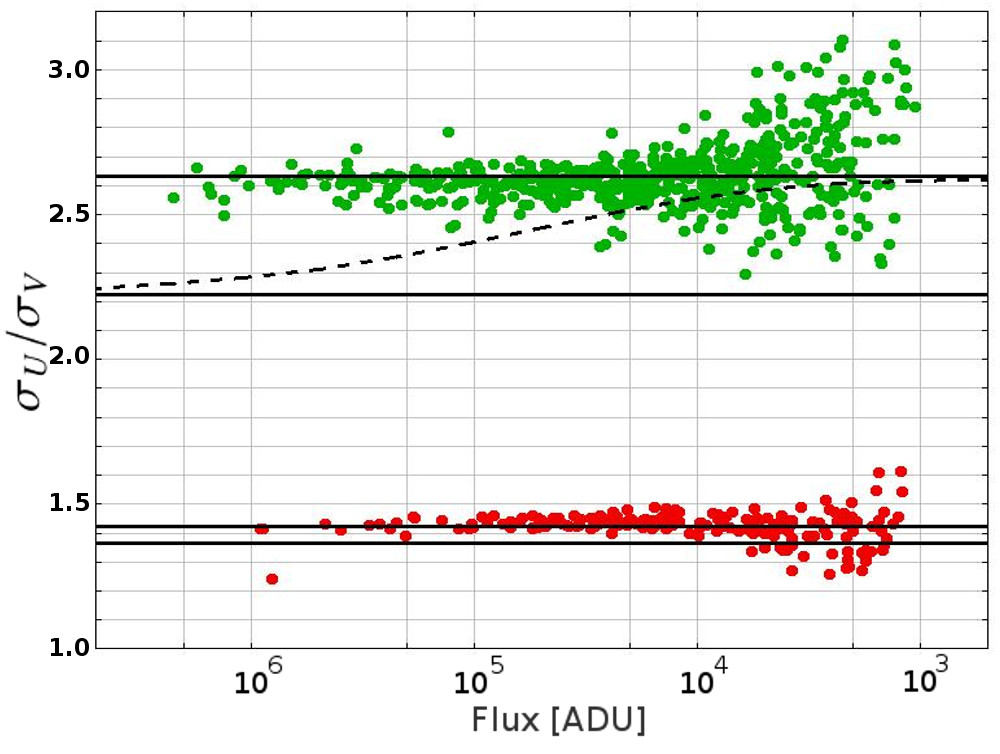}
\hspace{0.5cm}
\caption{$\sigma_U$ to $\sigma_V$ ratio versus flux (in ADU). The
  circles are for the stars in the VST images: upper circles are for
  stars in the image $VST_F$, while lower circles are for the stars
  in $VST_S$ (see text). The dashed curve corresponds to the ratio of
  the two exact expressions of the CRLB, Eqs.~(\ref{eq-mov-CR2D}) and
  (\ref{eq-mov-CR2D_v}), with the parameters corresponding to the image
  $VST_F$. The two upper horizontal lines correspond to the
  approximations of this ratio for the image $VST_F$ in the case of
  fast-moving sources (expression (\ref{ratioB_f}) for faint sources
  and expression (\ref{ratioF_f}) for bright sources). The two
  lower horizontal lines correspond to the approximations of this
  ratio for the image $VST_S$ in the case of slow-moving sources
  (expression (\ref{ratioB_s}) for faint sources and expression
  (\ref{ratioF_s}) for bright sources).}
\label{fig_CRapp2}
\end{center}
\end{figure}

Finally, we also note from Fig.~\ref{fig_CRapp1} that for both faint
and bright sources, the ratio of the two uncertainty estimates, $\sigma_U$
and $\sigma_V$, seems to be approximately constant for the drifting
sources, in very good agreement with the ratio of their respective
CRLB given by Eqs.~(\ref{eq-mov-CR2D}) and (\ref{eq-mov-CR2D_v}). This
is visually confirmed in the upper part of Fig.~\ref{fig_CRapp2},
where the $\sigma_U$ to $\sigma_V$ ratios are plotted (with circles) as a function of source flux for all the stars detected in the VST image. The dashed curve is the corresponding CRLB ratio.  In this figure, we clearly see that a large majority of the star uncertainty ratios are along the line of the corresponding CRLB
  ratio, with a better agreement in the regime of faint sources. We also note that for several faint source (less than $10\%$), there is an increase of the ratio values that we cannot currently explain.

With the aim of validating the CRLB approximations given in this
paper, we deduce the following simple expressions for the CRLB ratio
by using the simplified expressions given in
Sect.~\ref{sec:CR_moving_2D} for Eqs.~(\ref{eq-mov-CR2D}) and
(\ref{eq-mov-CR2D_v}). For fast-moving sources (as in the
case of the VST images of January 6, 2014), the CRLB ratio can be
approximated by the following two expressions:
\begin{eqnarray}
\label{ratioB_f}
\Frac{\sigma_U}{\sigma_V} &\simeq &
\sqrt{\overline{L}\left(\sqrt{\pi}-\frac{1}{\overline{L}}\right)}
\simeq 2.63 ~~~~~\mathrm{if}~~\tilde{F}/\tilde{B}<<1
\end{eqnarray}
\begin{eqnarray}
\label{ratioF_f}
\Frac{\sigma_U}{\sigma_V} &\simeq & 1.05 ~ \sqrt{\overline{L} }
\simeq 2.22 ~~~~~\mathrm{if}~~\tilde{F}/\tilde{B}>>1
.\end{eqnarray}

Expression (\ref{ratioB_f}) (valid for fast-moving faint sources) is based on the ratio of Eqs.~(\ref{eqLB3_2Dm}) and (\ref{eqLB3_2Ds}), while expression (\ref{ratioF_f}) (valid or fast-moving bright sources) is based on the ratio of Eqs.~(\ref{eqLF3_2Dm}) and (\ref{eqLF2_2Ds}). The numerical values are for the GBOT-VST image of January 6, 2014, for which $\overline{L}$ equals $4.45$. We call this image $VST_F$.

On the other hand, for slow-moving sources, the CRLB ratio can be approximated by the following two expressions:
\begin{eqnarray}
\label{ratioB_s}
\Frac{\sigma_U}{\sigma_V} &\simeq &
\sqrt{\frac{1+\frac{1}{2}\left(\overline{L}\right)^2+\frac{1}{12}\left(\overline{L}\right)^4}{1+\frac{1}{6}\left(\overline{L}\right)^2-\frac{1}{180}\left(\overline{L}\right)^4}}
\simeq 1.42 ~~~~\mathrm{if}~~\tilde{F}/\tilde{B}<<1
\end{eqnarray}
\begin{eqnarray}
\label{ratioF_s}
\Frac{\sigma_U}{\sigma_V} &\simeq &
\sqrt{1+\frac{1}{3}\left(\overline{L}\right)^2} \simeq 1.36
~~~~~~~~~~~~~~~~~~~\mathrm{if}~~\tilde{F}/\tilde{B}>>1
.\end{eqnarray}

Expression (\ref{ratioB_s}) (valid for slow-moving faint sources) is based on the ratio of Eq.~(\ref{eqLB2_2Dm}) and (\ref{eqLB2_2Ds}), while expression (\ref{ratioF_s}) (valid for slow-moving bright sources) is based on the ratio of Eqs.~(\ref{eqLF2_2Dm}) and
(\ref{eqLF2_2Ds}). As these expressions are for slow-moving sources, it is not possible to compare them with the error estimates of the VST image $VST_F$, where the star's drift is significant. Instead, for this aim, we selected another GBOT-VST image, taken on May 2, 2015, with a smaller drifting parameter $\overline{L}$ equal to $1.59$ (we call this second image $VST_S$). Then, we performed the astrometric reduction in a similar
manner as for the previous VST image, and we plot in Fig.~\ref{fig_CRapp2} the $\sigma_U$ to $\sigma_V$ ratios (with filled circles) for all its drifting stars. We also plot the two horizontal lines that correspond to expressions (\ref{ratioB_s}) and (\ref{ratioF_s}) for $VST_S$ (the numerical values given in the right part of these expressions are of course computed for the drifting parameter of this second image).

Figure~\ref{fig_CRapp2} shows that as expected, all the curves
and marks corresponding to the fast-drifting sources of the image
$VST_F$ are in the upper part of the diagram, largely above those
that correspond to the slow-moving sources of the image
$VST_S$. The smaller the drift, the smaller the
difference between the two uncertainty estimates (or the CRLB
values) along the $U$ and the $V$ axes, as extensively discussed in
the previous sections. For the fast-moving sources of $VST_F$ (upper
part of the figure), we see (as expected) that the exact expression of
the CRLB ratio (dashed curve) converges toward expression
(\ref{ratioF_f}) for bright sources and toward expression
(\ref{ratioB_f}) for faint sources (the two upper horizontal
lines). We did not plot the exact expression of the CRLB ratio for
the slow-moving sources of $VST_S$ to avoid crowding in the figure,
but in this case, it converges toward expression (\ref{ratioF_s})
for bright sources and toward expression (\ref{ratioB_s}) for faint
sources (the two lower horizontal lines), as should be the
case. This confirms that these approximate expressions for the CRLB in
the case of a two-dimensional array detectors are indeed correct.

Concerning the ratio of the uncertainty estimates along the $U$ and
$V$ axes, Fig.~\ref{fig_CRapp2} shows that for both fast- and slow-moving sources, they are in the areas defined by their respective CRLB
approximations (horizontal lines), but with a greater accumulation of
points along the upper horizontal lines that correspond to the CRLB
ratio approximation for faint sources. In fact, $90\%$ of the ratios of the error
estimates for all stars are in an interval of $10\%$ around
the CRLB ratio values computed under the faint source approximation.

We note that these last Eqs.~(\ref{ratioB_f}), (\ref{ratioF_f}),
(\ref{ratioB_s}), and (\ref{ratioF_s}) are particularly interesting
since they allow us to estimate the relative
difference of the centroiding precision quite accurately along the drifting direction and
along the direction normal to the motion, without knowing anything
except the source drifting parameter $\overline{L}$.

All these comparisons with both real and simulated data prove
the usefulness of the CRLB expressions presented in this paper in
characterizing the astrometric performance of moving sources
that are observed with digital sensors.

\section{Conclusions.}
\label{sec:Conclusion}
Following the earlier study of MSL13, we have extended their results
to include not only the more realistic case of a two-dimensional
detector array, but also explored the CRLB for moving objects. 
We have examined various different
plausible observational scenarios: over-, well-, and undersampled
pixel scales, and bright and faint objects (with respect to the background
level). In several interesting regimes we were able to simplify the
mathematics and obtain closed-form analytical equations for the CRLB,
which led to expressions that can be used to analyze data in a very
straightforward way. Thus we end up with a set of equations that
allows us to determine the ultimate limit of the astrometric quality
for astronomical data obtained with digital sensor arrays, today the
most common type of detectors in optical astronomy. These expressions
were validated by extensive comparisons with astrometric
reductions that were performed on several sets of both simulated and real
astronomical images. We note that the ultimate limit of the astrometric precision can be also deteriorated by several effects that we
did not examine in the current paper, such as the quality of the reference catalog that is used for the calibration or shutter-timing errors.

One of the interesting and new results of our analysis is that
  the maximum positional precision as predicted by the CRLB exhibits
  important oscillations when the pixel size falls between the image
  quality FWHM and the elongation of the source drift, probably because of
  the discretization of the source PSF by the detector array. For very small pixels or for severely undersampled images,
  these oscillations disappear. We provide a simple recipe to avoid
  these oscillations if the speed of the source is known (see
  Sect.~\ref{subsec:StudyCR_CSTBGperDEG}).

Another unexpected and very useful result of this study is that
it provides a simple method for computing an optimum exposure time
that minimizes the astrometric uncertainty of a drifting source
that is observed with one- and two-dimensional array detectors (see Sect. \ref{subsec:OptExpTime}). This method
should be particularly useful for planning astrometric observations of
asteroids, artificial satellites, space debris, and more generally for
any object drifting with respect to the focal plane. More detailed
comparisons of this theoretical result with astronomical observations
are necessary and will be performed in future studies.
  
The expressions developed in this study can be used to plan
observations, that is, to assess a priori the precision that
can be reached, to
maximize the yield of a set of observations, and to analyze the
achievable astrometric precision on existing data. In principle, they
could also be incorporated into data-simulation programs, such as
exposure-time calculators, which are widely used today
to plan
observations. This is especially beneficial in the age of large and
giant telescopes, whose operation is increasingly expensive, and which must
therefore be used in an optimum fashion.

Following the second study, \citet{Mendez14}, our results can also be
extended to assess the quality of the photometry of drifting
sources. Furthermore, our results could be adapted to other scenarios,
such as non-square pixels and non-circular PSFs that are due to optical
deficiencies of the instrumentation or the nature of the target
objects (e.g., galaxies). As explained in
Sect.~\ref{subsec:CompWithSimData}, we plan to incorporate more robust fitting algorithms in the GBOT pipeline, such as those based on
maximum likelihood, to avoid the loss of optimality of the LS
techniques that was recently described by \citet{Lobos15}.

\section{Acknowledgments.}
\label{sec:Acknowledgements}

\begin{small}
RAM and SB acknowledge support from CONICYT-FONDECYT grant No. 115 1213,
from project IC120009 ``Millennium Institute of Astrophysics (MAS)'' of
the Iniciativa Cient\'{\i}fica Milenio del Ministerio de
Econom\'{\i}a, Fomento y Tu\-ris\-mo de Chile, and from the Basal Center
for Astrophysics and Associated Technologies CATA PFB-06. RAM also
acknowledges ESO/Chile for hosting him during his sabbatical
leave
during 2014 in which part of this work was carried out. MA acknowledges support 
from the German Space Agency DLR on behalf of the German Ministry 
of Economy and Technology via grant 50 QG 1401. SB also acknowledges support from CNES in the frame of the CNES/INSU convention No. 151680. We have made extensive use of the  
Simbad database and the resources of the CDS in Strasbourg.
\end{small}

\begin{appendix}
\section{Calculation of integrals depending on the normalized PSF $\overline{\Phi}$}
 
\subsection{Integrals depending on $\overline{\Phi}_{S}$ and $\overline{\Phi}_{S^2}$} 
\label{app:S}

The three following integrals, which depend on $\overline{\Phi}_{S}$,
frequently appear in the CRLB expressions:
\begin{eqnarray}   
I_{S_1} &=& \frac{1}{\int_{-\infty}^{\infty}\left(\Derpar{\overline{\Phi}_{S}}{x_c}\right)^2
dx}= 4\sqrt{\pi}\sigma^3 \nonumber\\
I_{S_2} &=& \frac{1}{\int_{-\infty}^{\infty}\frac{1}{\overline{\Phi}_{S}} \left(\Derpar{\overline{\Phi}_{S}}{x_c}\right)^2 dx} = \sigma^2 \nonumber\\
I_{S_3} &=& \frac{1}{\int_{-\infty}^{\infty}\left(\overline{\Phi}_{S}\right)^2 dx}=2\sqrt{\pi}\sigma. \nonumber
\end{eqnarray}
The values of two integrals involved in expressions (\ref{eqLB-2D}) and (\ref{eqLF-2D}) when $\overline{\Phi}$ is replaced by $\overline{\Phi}_{S^2}$ can be deduced from $I_{S_1}$, $I_{S_2}$ and $I_{S_3}$:
\begin{eqnarray}
I_{S_1^2}= \frac{1}{\iint_{-\infty}^{\infty}\left(\Derpar{\overline{\Phi}_{S^2}}{x_c}\right)^2
dx dy}
=\frac{1}{\int_{-\infty}^{\infty}\left(\overline{\Phi}_{S}\right)^2 dy}
\frac{1}{\int_{-\infty}^{\infty}\left(\Derpar{\overline{\Phi}_{S}}{x_c}\right)^2
dx} = 8\pi\sigma^4 &&\nonumber   
\end{eqnarray}
\begin{eqnarray}   
I_{S_2^2}=\frac{1}{\iint_{-\infty}^{\infty}\frac{1}{\overline{\Phi}_{S^2}}\left(\Derpar{\overline{\Phi}_{S^2}}{x_c}\right)^2
dxdy} 
= \frac{1}{\int_{-\infty}^{\infty}\overline{\Phi}_{S} dy}
\frac{1}{\int_{-\infty}^{\infty}\frac{1}{\overline{\Phi}_{S}}
\left(\Derpar{\overline{\Phi}_{S}}{x_c}\right)^2 dx} = \sigma^2 && \nonumber
\end{eqnarray}

\subsection{Integrals depending on $\overline{\Phi}_M$ and $\overline{\Phi}_{M^2}$for a small drifting parameter}
\label{app:MsL}
The Taylor expansions of $\overline{\Phi}_M$ and $\Derpar{\overline{\Phi}_M}{x_c}$ in the vicinity of a drifting parameter equal to zero are given to order eight by the following equations (where $\overline{X}=\frac{\left(x-x_c\right)}{\sigma}$ and $\overline{L}=\frac{L}{2\sigma}$):\begin{eqnarray}
\overline{\Phi}_M = \frac{e^{-\frac{\overline{X}^2}{2}}}{\sqrt{2\pi } \sigma }
\left[ 1 - \frac{1}{6}\overline{L}^2\left(1-\overline{X}^2\right) + 
\frac{1}{120}\overline{L}^4 \left(3-6\overline{X}^2+\overline{X}^4\right)\right. 
~~~~~~~~ && \nonumber\\ 
- \left.\frac{1}{5040}\overline{L}^6
\left(15-45 \overline{X}^2+15\overline{X}^4-\overline{X}^6\right)\right]
+ o[\overline{L}^{8}]&&\nonumber
\end{eqnarray}
\begin{eqnarray}
\Derpar{\overline{\Phi}_M}{x_c}=
\frac{e^{-\frac{\overline{X}^2}{2}} \overline{X}}{\sqrt{2 \pi } \sigma^2}
\left[ 1 - \frac{1}{6}\overline{L}^2\left(3-\overline{X}^2\right)
+ \frac{1}{120}\overline{L}^4\left(15-10 \overline{X}^2+\overline{X}^4\right)
  \right. &&\nonumber\\  
- \left.\frac{1}{5040}\overline{L}^6\left(105-105 \overline{X}^2+21
    \overline{X}^4-\overline{X}^6\right)\right] + o[\overline{L}^{8}]&&\nonumber
\end{eqnarray}
These Taylor expansions allow us to approximate in the case of a small drifting parameter $L$, the three following integrals involved in the CRLB expressions of the moving sources:    
\begin{eqnarray}
I_{M_1}&=&\frac{1}{\int_{-\infty}^{\infty}\left(\Derpar{\overline{\Phi}_M}{x_c}\right)^2 dx} = 4\sqrt{\pi }\sigma^3 ~\left(1 + \frac{1}{2}\overline{L}^2 +
\frac{1}{12}\overline{L}^4\right) + o(\overline{L}^{8})\nonumber\\
I_{M_2}&=&\frac{1}{\int_{-\infty}^{\infty}\frac{1}{\overline{\Phi}_M}\left(\Derpar{\overline{\Phi}_M}{x_c}\right)^2 dx}
= \sigma^2 ~ \left(1 + \frac{1}{3}\overline{L}^2\right)
 + o(\overline{L}^{8})\nonumber\\
I_{M_3}&=&\frac{1}{\int_{-\infty}^{\infty}\left(\overline{\Phi}_M\right)^2 dx}=2\sqrt{\pi}\sigma\left(1 + \frac{1}{6}\overline{L}^2 -\frac{1}{180}\overline{L}^4\right) + o(\overline{L}^{6})\nonumber
\end{eqnarray}
For a small drifting parameter $L$, the values of the two integrals
involved in expressions (\ref{eqLB-2D}) and (\ref{eqLF-2D}) when
$\overline{\Phi}$ is replaced by $\overline{\Phi}_{M^2}$ can be
deduced from $I_{M_1}$, $I_{M_2}$, $I_{M_3}$ , and from $I_{S_1}$,
$I_{S_2}$ , and $I_{S_3}$ of Appendix~\ref{app:S}.  When the
derivative is performed with respect to $u_c$ (i.e., for the CRLB along
the drifting direction), the two integrals are written as follows:
\begin{equation}
\begin{array}{l}   
I_{M_1^2}^U=\Frac{1}{\iint_{-\infty}^{\infty}\left(\Derpar{\overline{\Phi}_{M^2}}{u_c}\right)^2 du dv}
=\Frac{1}{\int_{-\infty}^{\infty}\left(\overline{\Phi}_S\right)^2 dv}
~
\Frac{1}{\int_{-\infty}^{\infty}\left(\Derpar{\overline{\Phi}_M}{u_c}\right)^2
du}\\
~~~~~~~~~~~~~ = I_{S_3} ~ I_{M_1} = 8\pi\sigma^4 ~ \left(1 + \Frac{1}{2}\overline{L}^2 +\Frac{1}{12}\overline{L}^4\right) + o(\overline{L}^{8}) \\   
\\
I_{M_2^2}^U=\Frac{1}{\iint_{-\infty}^{\infty}\frac{1}{\overline{\Phi}_{M^2}}\left(\Derpar{\overline{\Phi}_{M^2}}{u_c}\right)^2 du dv} 
= \Frac{1}{\int_{-\infty}^{\infty}\overline{\Phi}_S dv}
~
\frac{1}{\int_{-\infty}^{\infty}\frac{1}{\overline{\Phi}_M}
\left(\Derpar{\overline{\Phi}_M}{u_c}\right)^2 du}\\
~~~~~~~~~~~~~~~~~~~~~~~~~~~~~~~~~~~~~=I_{M_2} = \sigma^2 ~ \left(1 + \Frac{1}{3}\overline{L}^2\right)
 + o(\overline{L}^{8})
\end{array}\nonumber
\end{equation}
When the derivative is performed with respect to $v_c$
(i.e., for the CRLB along the direction normal to the motion), the two
integrals are written as follows:
\begin{equation}
\begin{array}{l}   
I_{M_1^2}^V=\Frac{1}{\iint_{-\infty}^{\infty}\left(\Derpar{\overline{\Phi}_{M^2}}{v_c}\right)^2 du dv} = 
\Frac{1}{\int_{-\infty}^{\infty}\left(\Derpar{\overline{\Phi}_S}{v_c}\right)^2
dv} 
\Frac{1}{\int_{-\infty}^{\infty}\left(\overline{\Phi}_M\right)^2 du}\\
~~~~~~~~~~~~~ = I_{S_1} ~ I_{M_3} = 2\sqrt{\pi}\sigma\left(1 + \Frac{1}{6}\overline{L}^2 -\Frac{1}{180}\overline{L}^4\right) + o(\overline{L}^{6})\\
\\ 
I_{M_2^2}^V=\Frac{1}{\iint_{-\infty}^{\infty}\frac{1}{\overline{\Phi}_{M^2}}\left(\Derpar{\overline{\Phi}_{M^2}}{v_c}\right)^2 du dv} =
\Frac{1}{\int_{-\infty}^{\infty}\frac{1}{\overline{\Phi}_S} \left(\Derpar{\overline{\Phi}_S}{v_c}\right)^2 dv} 
\Frac{1}{\int_{-\infty}^{\infty}\overline{\Phi}_M du}\\
~~~~~~~~~~~~~~~~~~~~~~~~~~~~~~~~~~~~~~~ = I_{S_2} = \sigma^2
\end{array}\nonumber
\end{equation}

\subsection{Approximation of $\overline{\Phi}_M$, $\Derpar{\overline{\Phi}_M}{x_c}$ and related functions for a large drifting parameter}
\label{app:MbL}

For a large drifting parameter $L$, $\overline{\Phi}_M$ is a
quasi-constant function equal to $1/L$ except at each end of the
function (in the vicinity of $x =(x_c-L/2)$ and $x = (x_c+L/2)$). In a
similar way, the derivative of $\overline{\Phi}_M$ with respect to
$x_c$ (or $u_c$ or $v_c$) will be different from zero only in the
vicinity of these two areas.  Taking into account that the function
$\overline{\Phi}_M$ and its derivative are even functions, the
integrals of $I_{M_1}$, $I_{M_2}$ and $I_{M_3}$ (defined in
Appendix~\ref{app:MsL}) are well approximated for a large drifting
parameter by twice their integral over the interval around the
position $x = (x_c-L/2)$. We therefore first approximate the functions
$\overline{\Phi}_M$ and $\Derpar{\overline{\Phi}_M}{x_c}$ by the two
following functions by substituting $(x-x_c+L/2)$ with
$\sigma\cdot\gamma_x$ , where $\gamma_x$ is a small variation of x
(compared to $L$) in the vicinity of $(x_c-L/2)$:
\begin{eqnarray}
\overline{\Phi}_M &=& \frac{1}{2 L} \left( P \left(
\frac{\gamma_x}{\sqrt{2}} \right) + 1
\right)\nonumber\\
\Derpar{\overline{\Phi}_M}{x_c} &=& -\frac{1}{\sqrt{2\pi}\sigma L}
e^{-\frac{1}{2}\gamma_x^2}\nonumber
\end{eqnarray}
For large drifting parameter, the integrals of $I_{M_1}$, $I_{M_2}$ , and $I_{M_3}$ defined in  in Appendix~\ref{app:MsL} are then
well approximated by twice the integrals computed with the above approximations around $(x_c-L/2)$:
\begin{eqnarray}
I_{M_1} &\simeq& \pi\sigma L^2 ~ \frac{1}{\int_{-\infty}^{\infty}e^{-(\gamma_x)^2}
d\gamma_x} \simeq \sqrt{\pi}\sigma L^2\nonumber
\end{eqnarray}
\begin{eqnarray}
I_{M_2} &\simeq& \frac{\pi \sigma L}{2} ~ \frac{1}{\int_{-\infty}^{\infty}
\frac{e^{-\gamma_x^2}}{P\left(\frac{\gamma_x}{\sqrt{2}}\right) + 1} d\gamma_x} \simeq 0.55359 ~ \sigma L \nonumber\\
I_{M_3} &\simeq& \Frac{2 L^2}{\sigma} ~ \Frac{1}{\int_{-\infty}^{\frac{L}{2\sigma}}\left(P(\frac{\gamma_x}{\sqrt{2}})+1\right)^2 d\gamma_x} \simeq \frac{\sqrt{\pi}L^2}{\sqrt{\pi}L-2\sigma.}\nonumber
\end{eqnarray}
The agreement between these approximations and their exact expressions is
better than $2\%$, even for a drifting parameter $L=2$~FWHM.

For a large drifting parameter $L$, the integrals involved in functions $I_{M_1^2}^U$, $I_{M_2^2}^U$, $I_{M_1^2}^V$ , and $I_{M_2^2}^V$ defined in Appendix~\ref{app:MsL} become
\begin{eqnarray} 
I_{M_1^2}^U =& I_{S_3} ~ I_{M_1} &\simeq 2\pi\sigma^2L^2\nonumber\\ 
I_{M_2^2}^U =& I_{M_2} &\simeq 0.55359 ~ \sigma L \nonumber\\
I_{M_1^2}^V =& I_{S_1} ~ I_{M_3} &\simeq 4\pi\sigma^3\frac{L^2}{\sqrt{\pi}L-2\sigma}\nonumber\\ 
I_{M_2^2}^V =& I_{S_2} &\simeq \sigma^2\nonumber
\end{eqnarray}

\section{Calculation of the lower limit  $T_o$ of the optimum exposure time}
\label{app:AT_o}
As explained in Sect.~\ref{subsec:OptExpTime}, the lower limit $T_o$
of the optimum exposure time can be calculated as the root of the
derivative of expression~(\ref{eqLB}) with respect to exposure time
when $\overline{\Phi}$ is replaced by $\overline{\Phi}_M$.  With
Eq.~(\ref{eqBG2}) for the background, Eq.~(\ref{eqLB}) becomes
\begin{equation}
\sigma_{CR}^{2} = \left [ \frac{b_1 T_e + b_0}{f_s^2 T_e^2 \Delta    x}\right ]
~ \Frac{1}{\int_{-\infty}^{\infty}{\left(\Derpar{\overline{\Phi}_M}{x_c}\right)^2dx.}}\nonumber
\end{equation}

By denoting with $I$ the integral defined in the denominator of this expression, the derivative of $\sigma_{CR}^{2}$ with respect to exposure time is given by\begin{equation}
\Derpar{\sigma_{CR}^{2}}{T_e} = - \frac{b_1V_x^2}{f_s^2 \Delta x} ~ \left[\Frac{1}{L^2 I} + \Frac{1}{L I^2} ~ \Derpar{I}{L}\right]
-\frac{b_0 V_x^3}{f_s^2 \Delta x} ~ \left[\Frac{2}{L^3 I} +
  \Frac{1}{L^2 I^2} ~ \Derpar{I}{L}\right]\nonumber
\end{equation}

To find the lower limit of the optimum exposure time, we then
need to
solve
$$\left[1 + \Frac{L}{I} ~ \Derpar{I}{L}\right] + \gamma_b
\left[\Frac{2\sigma}{L} + \Frac{\sigma}{I} ~ \Derpar{I}{L}\right]
=0,$$ where $\gamma_b$ equals $\frac{b_0V_x}{b_1\sigma}$. The integral
$I$ can be expressed in terms of $\overline{L}$ (equal $to
  \frac{L}{2\sigma}$), and therefore we have
$$\left[1 + \gamma_b \overline{L} + 2\overline{L}^2 - e^{\overline{L}^2}\right] = 0.$$
A solution $\overline{L}_o$ of this equation exists as a power series of $\gamma_b$ converging for a low value of $\gamma_b$ (note that $\gamma_b$ will be low if we assume that the RON of the CCD is smaller than the sky background when the exposure time is close to $T_o$), 
\begin{eqnarray}
&&\overline{L}_o =  1.12 + 0.33\gamma_b - 0.24\gamma_b^2 + 0.25\gamma_b^3 +o(\gamma_b^4),
\nonumber
\end{eqnarray}
and then the corresponding drifting parameter $L_o$ according to image quality FWHM is \begin{equation}
L_o=\overline{L}_o\frac{\text{FWHM}}{\sqrt{2\ln{(2)}}} = \left[ 0.95 + 0.66\mu_b - 1.12\mu_b^2 + 2.75\mu_b^3 \right]~ \text{FWHM,} \nonumber
\end{equation}
where $\mu_b$ equals $\frac{b_0}{b_1 ~ T_s}$ (with $T_s$ the exposure time corresponding to a drifting elongation equal to
the FWHM). Then, $T_o$ is equal to $L_o/V_x$ , and its final expression is
\begin{equation}
T_o = \left[ 0.95 + 0.66\mu_b - 1.12\mu_b^2 + 2.75\mu_b^3\right] ~ T_s,
\end{equation}
where $T_s$ is $\text{the FWHM}/V_x$.

\end{appendix}

\end{document}